\documentclass[twocolumn,floatfix,superscriptaddress,a4paper,showpacs,showkeys,nofootinbib,reprint,prc,noeprint]{revtex4-1}
\usepackage[english]{babel}
\usepackage[utf8]{inputenc}
\usepackage[T1]{fontenc}

\usepackage[colorlinks=true,linktocpage=true,linkcolor=blue,citecolor=blue,allcolors=blue]{hyperref}
\usepackage{url}

\usepackage{epsfig}
\usepackage{float}

\usepackage{amsmath, amssymb, physics}

\usepackage{enumerate, color, xcolor, latexsym, xfrac, needspace}

\usepackage{longtable}
\usepackage{booktabs}
\usepackage{rotating}

\begin{document}

\title{$K^*(892)$ Resonance Suppression in Ar+Sc Collisions at SPS Energies}

\author{Amine Chabane}
\affiliation{Institut f\"{u}r Theoretische Physik, Goethe-Universit\"{a}t Frankfurt, Max-von-Laue-Str. 1, D-60438 Frankfurt am Main, Germany}
\affiliation{Helmholtz Research Academy Hesse for FAIR (HFHF), GSI Helmholtzzentrum f\"ur Schwerionenforschung GmbH, Campus Frankfurt, Max-von-Laue-Str. 12, 60438 Frankfurt am Main, Germany}

\author{Tom Reichert}
\affiliation{Institut f\"{u}r Theoretische Physik, Goethe-Universit\"{a}t Frankfurt, Max-von-Laue-Str. 1, D-60438 Frankfurt am Main, Germany}
\affiliation{Frankfurt Institute for Advanced Studies (FIAS), Ruth-Moufang-Str. 1, D-60438 Frankfurt am Main, Germany}
\affiliation{Helmholtz Research Academy Hesse for FAIR (HFHF), GSI Helmholtzzentrum f\"ur Schwerionenforschung GmbH, Campus Frankfurt, Max-von-Laue-Str. 12, 60438 Frankfurt am Main, Germany}

\author{Jan Steinheimer}
\affiliation{GSI Helmholtzzentrum f\"ur Schwerionenforschung GmbH, Planckstr. 1, 64291 Darmstadt, Germany}
\affiliation{Frankfurt Institute for Advanced Studies (FIAS), Ruth-Moufang-Str. 1, D-60438 Frankfurt am Main, Germany}

\author{Marcus Bleicher}
\affiliation{Institut f\"{u}r Theoretische Physik, Goethe-Universit\"{a}t Frankfurt, Max-von-Laue-Str. 1, D-60438 Frankfurt am Main, Germany}
\affiliation{Helmholtz Research Academy Hesse for FAIR (HFHF), GSI Helmholtzzentrum f\"ur Schwerionenforschung GmbH, Campus Frankfurt, Max-von-Laue-Str. 12, 60438 Frankfurt am Main, Germany}

\date{\today}

\begin{abstract}

We investigate the production and suppression of short-lived $K^*(892)$ resonances in p+p and Ar+Sc collisions at CERN-SPS energies ($\sqrt{s_{\mathrm{NN}}} = $ 8.8, 11.9, and 16.8~GeV) using the Ultra-relativistic Quantum Molecular Dynamics (UrQMD) model. We present multiplicities, rapidity and transverse momentum distributions, and analyze the $K^*/K$ yield ratios as a function of energy and centrality. We further estimate the time interval between chemical and kinetic freeze-out using the experimental method. A detailed comparison with recent NA61/SHINE data demonstrates that the UrQMD model captures the essential features of resonance dynamics, although the very strong resonance suppression in central collisions observed in the data cannot be quantitatively reproduced.
\end{abstract}

\maketitle

\section{Introduction}
The physics of strong interaction as described by Quantum Chromodynamics (QCD) is nowadays under investigations at the worlds largest accelerator facilities, namely at CERN (running programs at the SPS and the LHC), at Brookhaven National Laboratory using the RHIC to perform collider and fixed target experiments or at GSI's  Schwerionensynchrotron (SIS18) and with new facilities planned at FAIR (especially with the CBM-experiment), HIAF and NICA.
The special focus of these programs is to explore the properties of hot and/or dense QCD matter, i.e. its bulk properties, expansion dynamics and thermodynamic properties \cite{Sorensen:2023zkk}. 

Lattice QCD calculations suggest that in the center-of-mass energy range between 4 GeV and 30 GeV a transition to a deconfined state of quarks and gluons might happen. It is also speculated that a change from a cross over to a first order phase transition might happen in this range which would imply the existence of a critical end point in the QCD phase diagram. Unfortunately, clear experimental signatures of the transition to the deconfined state and for the critical end point are still not observed, although circumstantial evidence has been reported. A main signal for a first order phase transition would be an enhanced life-time of the fireball, due to a softening of the equation-of-state caused by a large latent heat. Such an enhanced life-time can e.g. be seen in Hanbury-Brown-Twiss interferometry (HBT) \cite{HanburyBrown:1954amm,Goldhaber:1960sf,Bauer:1992ffu,Pratt:1986cc} which uses quantum correlations of e.g. pion-pion or proton-proton pairs to constrain the size and duration of the system \cite{STAR:2014shf,PHENIX:2014dmi,Armesto:2015ioy,HADES:2019lek}. Lifetime estimates using HBT are sensitive to a phase transition \cite{Li:2022iil}.

An alternative approach to estimate the life-time of the (hadronic part) of the fireball is given by the suppression of hadronic resonances. Let us assume that one can divide the systems evolution into different stages, a chemical and kinetic freeze-out regime. Then, during the expansion, the scattering rates compete with the expansion rate until the so-called kinetic freeze-out is reached and the system decouples. During the time span from chemical to kinetic freeze-out hadron resonances are produced and decay. Thus, the final yield of reconstructable resonances is usually suppressed \cite{Torrieri:2001ue,Rafelski:2001hp,Bleicher:2002dm,Markert:2002rw,Knospe:2015rja,Knospe:2015nva,Ilner:2016xqr,Oliinychenko:2021enj,Sahoo:2023rko,Sahoo:2023dkv} as compared to the yields present at the chemical freeze-out hypersurface. This suppression is due to the rescattering of the decay daughter particles from the resonance that undergo rescattering and shift the pair invariant mass out of the resonance peak.

To be more specific, most hadronic resonances have lifetimes on the order of 1-5~fm/c (like the $\rho$, $\Delta$ and the $K^*$), with a few exceptions like the $\phi$ living approx. 50~fm/c. These short lived states decay inside the system while it is expanding. The most commonly measured resonance explored for this approach is the $K^*(892)$ resonance \cite{STAR:2004bgh,Markert:2005jv,STAR:2008twt,HADES:2013sfy,ALICE:2018ewo,ALICE:2022zuc,ALICE:2023ifn}, having a vacuum lifetime of approx. 4 fm/c. This has also been extensively investigated on the theory side \cite{Bleicher:2002dm,Knospe:2015rja,Knospe:2015nva,Ilner:2016xqr,Knospe:2021jgt,Chabane:2024crn,Neidig:2025xgr}

Recently, the NA61/SHINE collaboration at CERN has reported measurements on the $K^*/K$ ratio in intermediate mass Ar+Sc reactions \cite{Kozlowski:2024cjw}. These measurements complement existing measurements by STAR \cite{STAR:2022sir} at similar energies and allow to study the system size dependence of the lifetime of a heavy-ion collision.

Following our investigations of the STAR@RHIC data for Au+Au reactions we explore here the $K^*(892)$ resonance suppression and subsequent lifetime estimates in Ar+Sc collisions and compare to the NA61/SHINE data. To this aim we will use the Ultra-relativistic Quantum Molecular Dynamics transport model (v3.6).

\begin{figure} [t!]
    \centering
    \includegraphics[width=\columnwidth]{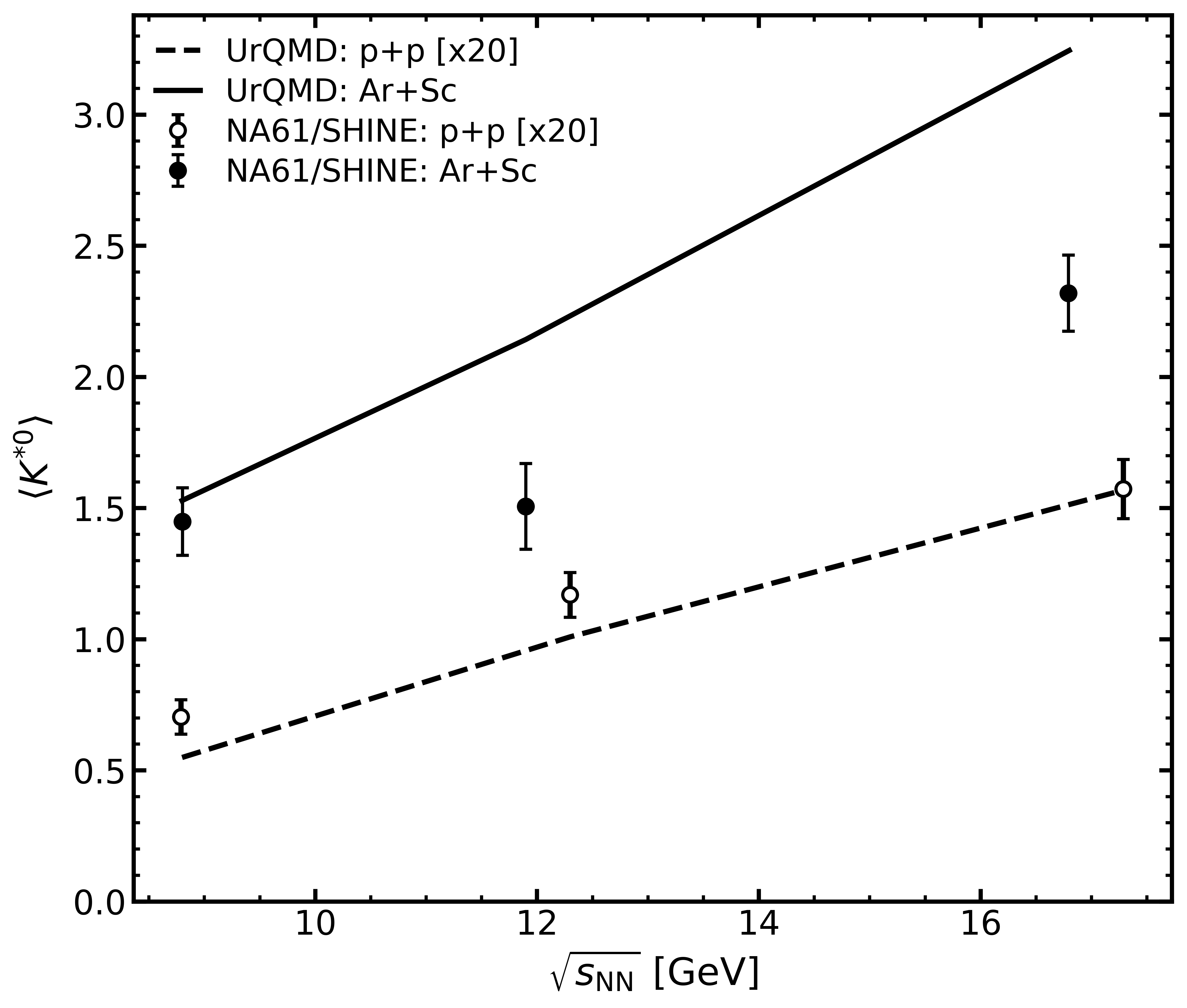}
    \caption{Average multiplicity of $K^{*0}(892)$ resonances as a function of the center-of-mass energy $\sqrt{s_{NN}}$ for p+p and central (0-10$\%$) Ar+Sc collisions. The dashed line shows UrQMD model calculations for p+p collisions (scaled by a factor of 20), while the solid line represents UrQMD predictions for Ar+Sc collisions. Open circles indicate NA61/SHINE experimental data for p+p collisions (scaled by a factor of 20), and filled circles show NA61/SHINE data for Ar+Sc collisions \cite{Kozlowski:2024cjw}. Error bars represent statistical uncertainties.}
    \label{fig:plot_sNN_mult}
\end{figure}

\section{Model and Methods}

\subsection{Ultra-relativistic Quantum Molecular Dynamics}

The Ultra-relativistic Quantum Molecular Dynamics (UrQMD) model \cite{Bass:1998ca,Bleicher:1999xi,Bleicher:2022kcu} is a relativistic transport model of QMD type \cite{Aichelin:1986wa,Aichelin:1991xy}. The degrees of freedom of the model are hadrons (currently approx. 100 different hadrons and their resonances are implemented in accordance with the PDG \cite{ParticleDataGroup:2024cfk}) which are covariantly propagated. Binary interactions are governed by the geometrical interpretations of (in-)elastic scattering cross sections, which are taken from experimental data if available or derived from effective models. UrQMD allows for string excitation, following the Lund string picture, and subsequent fragmentation. The dynamics of resonances is mainly governed by the production and regeneration in meson-meson and meson-baryon interactions. The cross sections are fitted to available experimental data or extrapolated in given reaction classes \cite{Bass:1998ca,Bleicher:1999xi}. The decay branching ratios are taken from the PDG and by the corresponding Clebsch-Gordon coefficients. The formation and decay of resonances are coupled via detailed balance relations.

\subsection{Identification of observable resonances}
Typically, hadronic resonances have two distinct decay branches: I) electro-magnetic decays, e.g $\rho \rightarrow e^+e^-$ or II) hadronic decay channels, e.g. $K^* \rightarrow K\pi$. Some (vector) resonance like the $\rho$, $\phi$ or $J/\psi$ can even be observed in both channels. Due to experimental difficulties most experiments, however have only access to the hadronic decay branch. Here, a resonance is measured indirectly from an invariant mass analysis of  its decay products. The invariant mass analysis of all daughter particle pairs has to be separated from the uncorrelated background and then the resonance signal is extracted by fitting a relativistic Breit-Wigner distribution to the remaining signal in the region of the resonance's pole mass in vacuum. Such an approach is statistically demanding, as the background needs to be estimated with high statistics to allow for a meaningful extraction of the signal. 

Theoretically, one can circumvent this problem (and also obtain more detailed information) by using an alternative method. In the simulation, we track all decay products of each resonance through the future evolution of the system and only in the case that none of the decay daughter scatters we consider the resonance as reconstructable. This method is well established and has been employed in several studies \cite{Bleicher:2002dm,Vogel:2009kg, Knospe:2015rja,Knospe:2015nva,Steinheimer:2015sha,Ilner:2017tab,Reichert:2019lny}. It has also been checked that it provides the same results as a full fledged invariant mass analysis \cite{Chabane:2024crn}.

\section{Results}

We analyze Ar+Sc collisions at center-of-mass energies of $\sqrt{s_\mathrm{NN}} = $ 8.8, 11.9 and 16.8 GeV as measured by the NA61/SHINE collaboration. The centrality is estimated using the impact parameter (here 0-10\% central events are simulated using a geometrically weighted impact parameter range of $b=0-2.67$ fm). The number of participants is calculated for each simulation in the Glauber approximation. We also show p+p collisions at the same center-of-mass energies of $\sqrt{s_\mathrm{NN}} = $ 8.8, 11.9 and 16.8 GeV. The analysis focuses on the $K^{*0}(892)$ resonance which is experimentally reconstructed in the charged $\pi^- + K^+$ channel. For the sake of brevity, we will denote the $K^{*0}(892)$ simply as $K^*$.

In Fig. \ref{fig:plot_sNN_mult} we show the average multiplicity of $K^{*0}(892)$ resonances as a function of the center-of-mass energy $\sqrt{s_{NN}}$ for p+p and central (0-10$\%$) Ar+Sc collisions. The dashed line shows UrQMD model calculations for p+p collisions (scaled by a factor of 20), while the solid line represents UrQMD predictions for Ar+Sc collisions. Open circles indicate NA61/SHINE experimental data for p+p collisions (scaled by a factor of 20), and filled circles show NA61/SHINE data for Ar+Sc collisions. We clearly observe that the model calculations describe the experimental data on Kaon resonance production in p+p collisions in this energy range very well. The $K^*$ yields in central Ar+Sc are overestimated in UrQMD for the two highest beam energies.


\begin{figure} [t!]
    \centering 
    \includegraphics[width=\columnwidth]{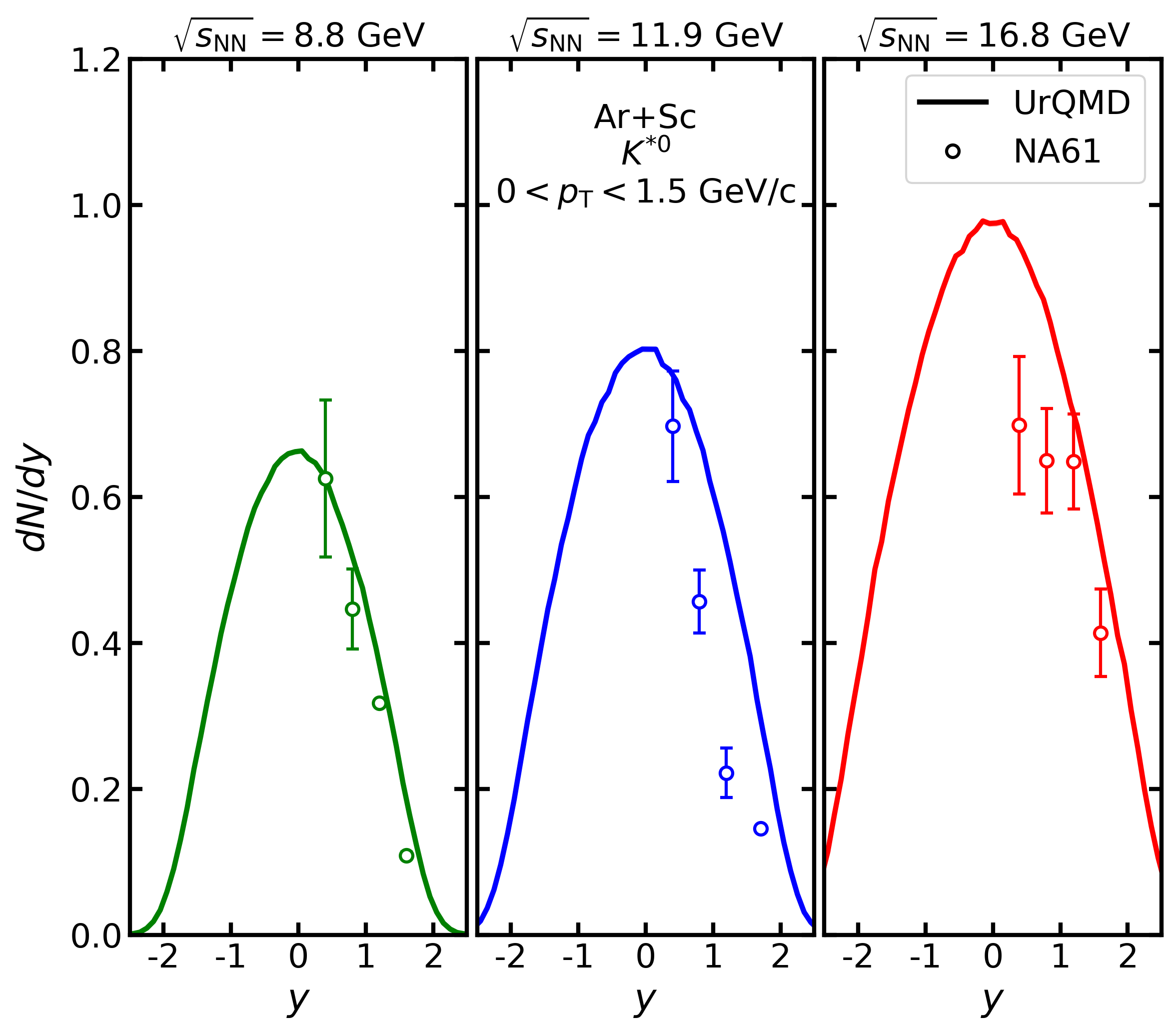}
    \caption{ Rapidity distributions ($dN/dy$) of reconstructable $K^{*0}(892)$ resonances in the transverse momentum range $0 \leq p_\mathrm{T} \leq 1.5$~GeV in 10\% central Ar+Sc collisions at $\sqrt{s_\mathrm{NN}} =$ 8.8 (left, green), 11.9 (middle, blue), and 16.8 (right, red)~GeV. The UrQMD model calculations are shown as solid lines, whereas the NA61/SHINE data \cite{Kozlowski:2024cjw} are shown as symbols.}
    \label{fig:plot_y_dNdy_all}
\end{figure}

\subsection{Rapidity and transverse momentum distributions}

We continue the investigation by comparing the rapidity distributions and the transverse momentum distributions of the reconstructable $K^{*0}(892)$ resonances. In Fig.~\ref{fig:plot_y_dNdy_all} the rapidity distributions are shown for 10\% central Ar+Sc collisions at $\sqrt{s_\mathrm{NN}} =$ 8.8 (left, green), 11.9 (middle, blue), and 16.8 (right, red)~GeV. The UrQMD model calculations are shown as solid lines, whereas the NA61/SHINE data \cite{Kozlowski:2024cjw} are shown as full symbols. As in the data, the transverse momentum range in the simulation is restricted to $0 \leq p_\mathrm{T} \leq 1.5$~GeV. The calculations agree generally well with the measured data by NA61/SHINE. However, at the highest energy, the suppression in the data at midrapidity seems to be stronger than observed in the simulation, however this deviation is essentially based on one data point.

\begin{figure} [t!]
    \centering
    \includegraphics[width=\columnwidth]{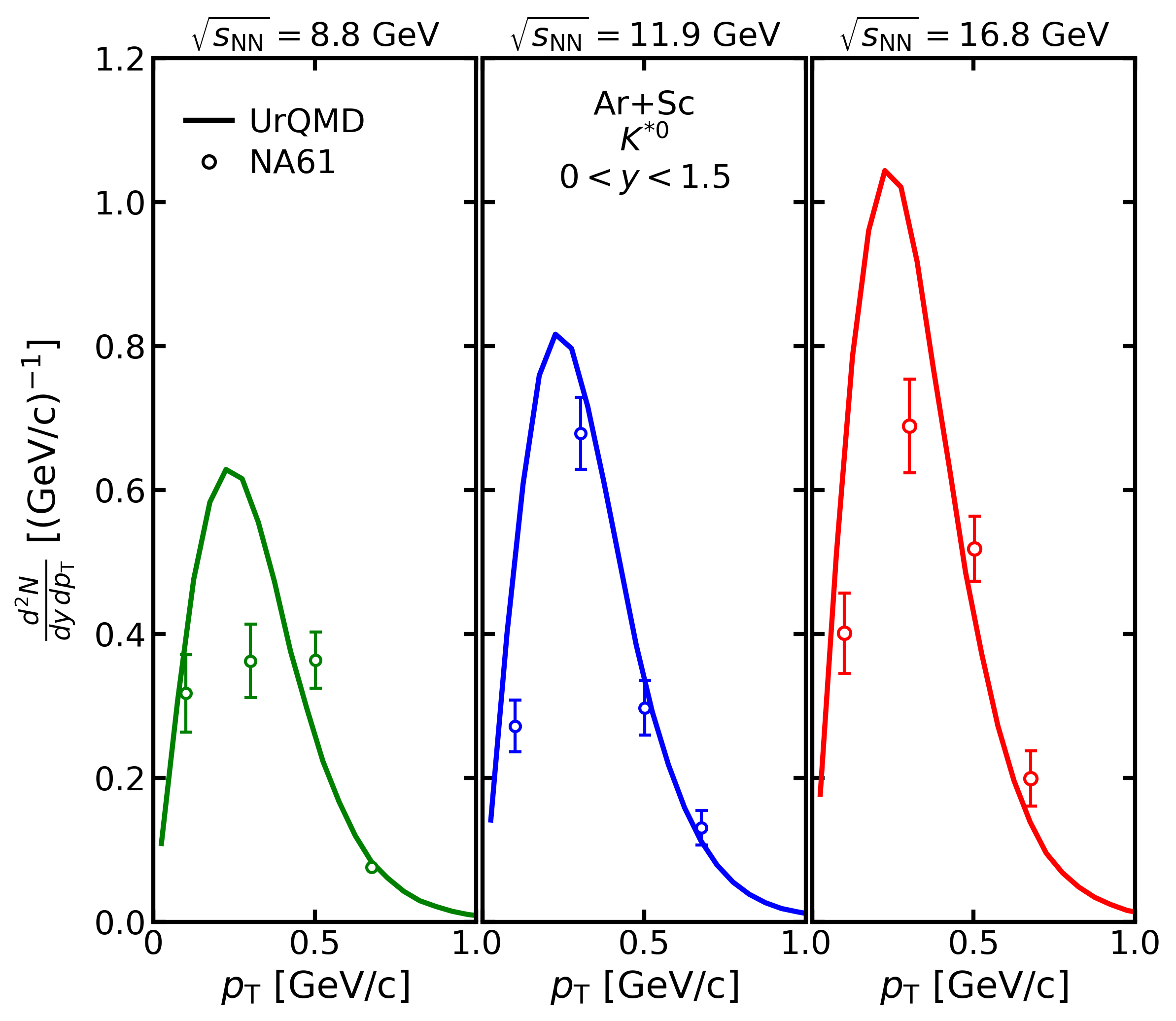}
    \caption{Transverse momentum distributions ($dN/(dy dp_\mathrm{T})$) of reconstructable $K^{*0}(892)$ resonances in the rapidity range $0 \leq y \leq 1.5$ in 10\% central Ar+Sc collisions at $\sqrt{s_\mathrm{NN}} = $ 8.8 (left, green), 11.9 (middle, blue), and 16.8 (right, red)~GeV. The UrQMD model calculations are shown as solid lines, whereas the NA61/SHINE data \cite{Kozlowski:2024cjw} are shown as symbols.}
    \label{fig:plot_pT_dNdpT}
\end{figure}

Next, Fig. ~\ref{fig:plot_pT_dNdpT} presents the transverse momentum distributions ($dN/(dy dp_\mathrm{T})$) of reconstructable $K^{*0}(892)$ resonances in the rapidity range $0 \leq y \leq 1.5$ in 10\% central Ar+Sc collisions at $\sqrt{s_\mathrm{NN}} = $ 8.8 (left, green), 11.9 (middle, blue), and 16.8 (right, red)~GeV. The UrQMD model calculations are shown as solid lines, whereas the NA61/SHINE data \cite{Kozlowski:2024cjw} are shown as symbols.

The model calculations again agree well with the measured data by NA61/SHINE for most transverse momenta. At the lowest energy the experimental data appears to be rather flat at low transverse momenta, this is not observed in the simulations. However, the shape of the distribution at this energy is rather peculiar and can not be captured with any standard (thermally motivated) distribution. At the higher energies the $K^*$ yield around the maximum of the distribution is systematically overestimated by the model which leads to the deviation in the integrated yield. However, also the error bars should be taken into account and the preliminary nature of the results.

\begin{figure} [t]
    \centering
\includegraphics[width=\columnwidth]{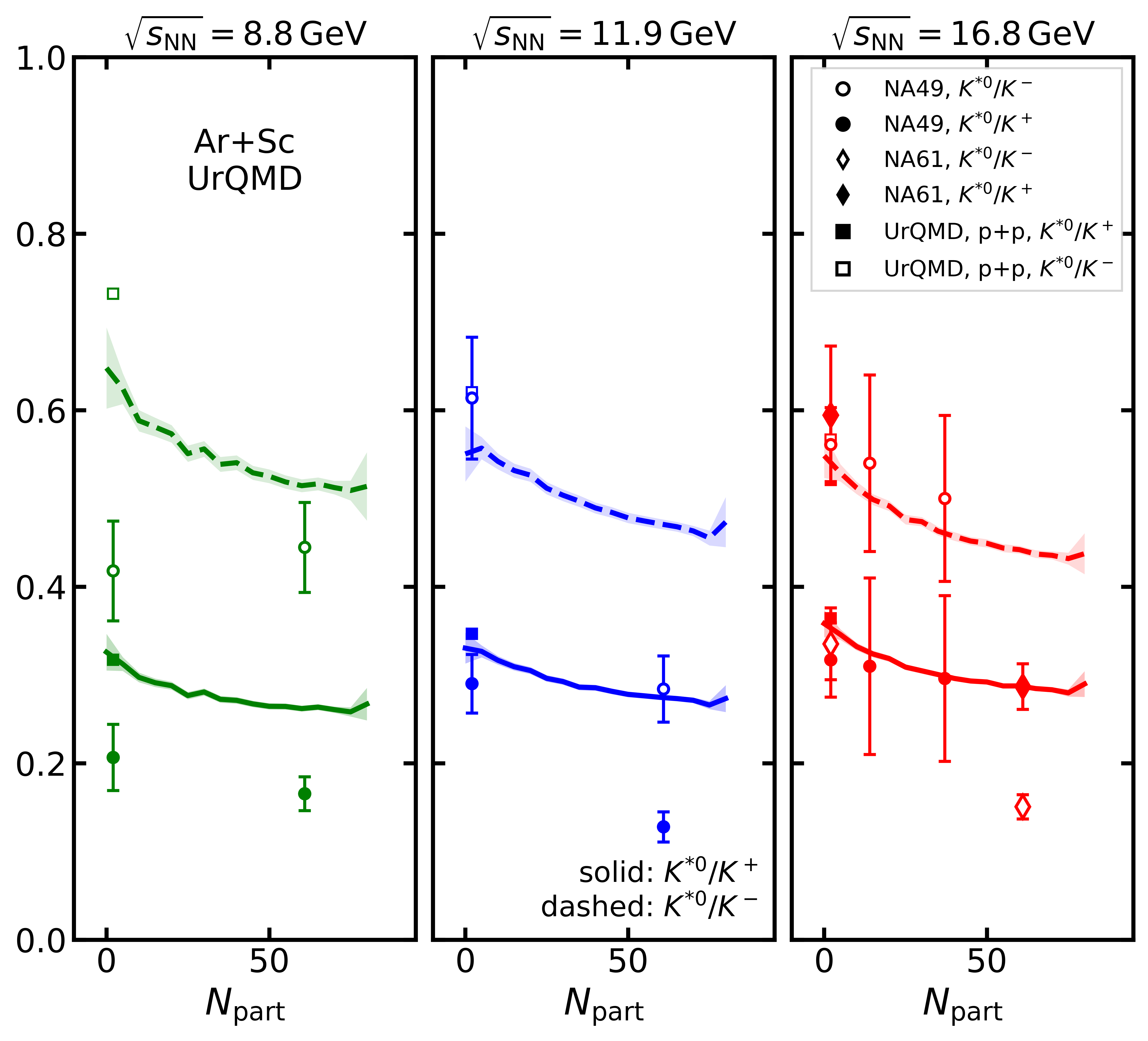}
    \caption{ Ratio of reconstructable $K^{*0}(892)/K^+$ (full lines, full symbols) and $K^{*0}(892)/K^-$ (dashed lines, open symbols) as function of $N_\mathrm{part}$ in Ar+Sc collisions at $\sqrt{s_\mathrm{NN}} = $ 8.8 (left), 11.9 (middle) and 16.8 (right)~GeV. The lines represent the UrQMD model calculations, while the symbols show to measurements from NA61/SHINE \cite{Kozlowski:2024cjw}.}
    \label{fig:plot_npart_ratio}
\end{figure}

\begin{figure} [t]
    \centering
    \includegraphics[width=\columnwidth]{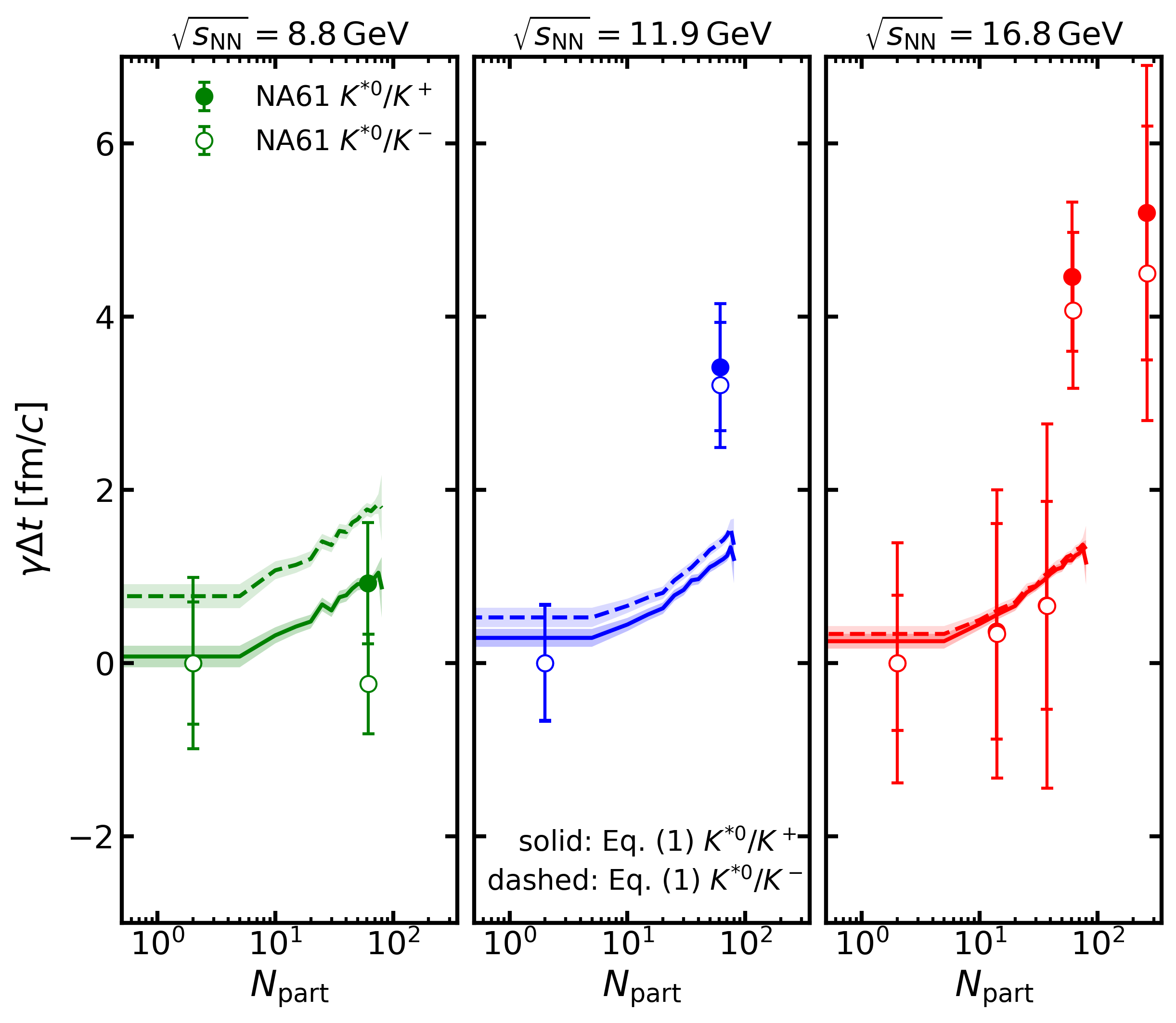}
    \caption{ Duration of the hadronic stage $\gamma\Delta t$ estimated as the difference between kinetic and chemical freeze-out as a function of $N_\mathrm{part}$ in Ar+Sc collisions at $\sqrt{s_\mathrm{NN}} = $ 8.8~GeV (left), 11.9~GeV (middle) and 16.8~GeV (right). The estimates are based on Eq. \ref{eq:dt_standard}. UrQMD calculations are shown as full lines (dashed lines) for the ratio $K^{*0}/K^+$ ($K^{*0}/K^-$), while full symbols (open symbols) show experimental $K^{*0}/K^+$ ($K^{*0}/K^-$) ratios converted to $\gamma\Delta t$ using Eq.~\ref{eq:dt_standard}. }
    \label{fig:plot_t_deltat_np}
\end{figure}

\subsection{Centrality dependence and the $K^{*0}/ K$ ratio}

Next, we investigate the centrality dependence of the reconstructable $K^{*0}(892)$ multiplicity. The centrality dependence of the $K^*/K$ ratio is depicted in Fig.~\ref{fig:plot_npart_ratio}. Here different ratios are explored, the ratio of $K^{*0}(892)/K^+$ (UrQMD: full line, data: full symbols) and $K^{*0}(892)/K^-$ (UrQMD: dashed line, data: open symbols) as function of $N_\mathrm{part}$ in Ar+Sc collisions at $\sqrt{s_\mathrm{NN}} = $ 8.8 (left), 11.9 (middle) and 16.8 (right)~GeV. The p+p results of the simulation are shown as squares.

It is evident that the $K^{*}/K^+$ ($K^{*}/K^-$) ratios decrease with increasing $N_\mathrm{part}$, a trend that can be generally attributed to the enhanced rescattering of the decay daughters in larger, more central collisions. This suppression effect is clearly captured by the UrQMD model and is in good agreement with the experimental observations. This suggests that the interpretation that the hadronic phase plays a significant role in diminishing the reconstructable resonance yield is generally valid. However, the current data also shows a surprisingly strong suppression that develops at the mid and highest energy for the most central reactions in the $K^*/K^-$ ratio. Such an abrupt drop may indicate an additional suppression mechanism not present in the current hadronic transport simulation. It may indicate e.g. an extend lifetime of the hadronic stage, e.g. due to the crossing of the first order phase transition.

Let us shortly comment on the difference of the $K^*/K^-$ ratio between peripheral Ar+Sc events and proton+proton reactions at the lowest investigated energy, $\sqrt{s_\mathrm{NN}}=8.8$ GeV. Here the nucleus-nucleus collisions are still a superposition of p+p, p+n and n+n reactions. The $K^-$ production is relatively suppressed in p+p reactions as it is produced on the diquark side of the string end. This often forces the associated production of a $\Delta^{++}$ at the diquark string, requiring a higher string mass, which results in a suppression of the $K^-$ as compared to n+n or n+p reactions which make up the peripheral nucleus-nucleus collision. Thus, it is natural to expect that the $K^*/K^-$ ratio in p+p is higher than in peripheral nucleus-nucleus reactions. One should note that this effect is usually not important for the $K^+$ production as the $K^+$ is mostly produced at the single quark end of the string. 

\subsection{Lifetime estimates}

Finally, we present an analysis of the duration of the hadronic state based on the resonance to ground state ratios in Ar+Sc collisions at SPS energies. Based on the assumption that the fireball evolution is characterized by two subsequent stages: I) The chemical freeze-out, where in-elastic, flavor changing reactions cease and primordial particle multiplicities (including those of resonances) are established. II) This stage is followed by the (pseudo-)elastic hadronic rescattering stage until the so-called kinetic freeze-out happens and all strong interactions cease. Assuming that the resonance multiplicity follows an exponential decay law (i.e. neglecting regeneration effects) and that all resonances that decay during the hadronic rescattering stage are not reconstructable one can relate the ratios $K^*/K$ at chemical and kinetic freeze-out to the lifetime of the hadronic stage $\Delta t \equiv t_\mathrm{kin.fo} - t_\mathrm{chem.fo}$:
\begin{align}
    \Delta t &\approx \tau_{K^*} \left[ \ln (K^*/{K})|_\mathrm{chem.fo} - \ln(K^*/{K})|_\mathrm{kin.fo} \right],
\label{eq:dt_standard}
\end{align}
where $\tau_{K^*}$ is the vacuum lifetime of the $K^*$. This estimate provides a lower limit on the duration of the hadronic stage, because it neglects regeneration. 

To obtain the values needed for Eq. \ref{eq:dt_standard},  the experiment reconstructs $K^{\ast}(892)$ from identified $K\pi$ pairs at the end of the reaction to obtain the ratios at kinetic freeze-out. To estimate the ''experimental'' ratios at chemical freeze-out, the ratios are assumed to be the $p+p$ values at the same energy. In our calculation we do exactly the same to allow for a comparison with the data.

It is possible to improve these estimates and to include the regeneration effects to obtain a more realistic estimate of the lifetime. Such an improved lifetime estimate was presented in \cite{Neidig:2021bal,Neidig:2025xgr}, where a full set of rate equations was used to describe the hadron-chemical reactions of the system from chemical to kinetic freeze-out. In such an approach the $K^*/K$ ratio falls off nearly linearly (instead of exponentially) when regeneration is taken into account\footnote{%
To connect both estimates, one can Taylor expand the logarithm in Eq. \ref{eq:dt_standard} around the typical $(K^* / K)\approx0.5$ value ($\ln(x)|_{x=0.5}\approx -\ln(2)+2(x-0.5)$), giving
\begin{align*}
    \Delta t &\approx \tau_{K^*} \left[ \ln (K^*/{K})|_\mathrm{chem.fo} - \ln(K^*/{K})|_\mathrm{kin.fo} \right]\\
&\approx 2\tau_{K^*} \left[(K^*/{K})|_\mathrm{chem.fo} - (K^*/{K})|_\mathrm{kin.fo}\right]
\end{align*}
Using the lifetime $t_{K^*}\approx 4$fm/c and comparing to $T=1/\Gamma=50$ fm/c this indicates that a substantial amount of $K^*$ are regenerated during the evolution from chemical to kinetic freeze-out.}. %
This approach leads to an estimate for the lifetime of the hadronic system given by 
\begin{align}
    \Delta t &\approx T\left[(K^* / K)|_\mathrm{chem.fo} -  (K^* / K)|_\mathrm{kin.fo}\right],
\end{align}
where $T=1/\Gamma$, with $\Gamma = 0.02\,\mathrm{fm}^{-1}$ being an essentially energy independent damping rate (for nucleus-nucleus collisions in the energy range from $\sqrt{s_\mathrm{NN}}=7.7-39$ GeV) at which the ratio decreases with time \cite{Neidig:2025xgr}.

To compare to the experimental data, we also include a Lorentz-gamma factor (as done in the NA61 analysis) given by
$\gamma = \sqrt{1+(\langle p_T\rangle_{K^*}/m_{K^*})^2}$, where the mean transverse momenta are on the order of 600-650 MeV, depending on collision energy and centrality. 

Fig. ~\ref{fig:plot_t_deltat_np} shows the duration of the hadronic stage $\gamma\Delta t$ estimated as the difference between kinetic and chemical freeze-out as a function of $N_\mathrm{part}$ in Ar+Sc collisions at $\sqrt{s_\mathrm{NN}} = $ 8.8~GeV (left), 11.9~GeV (middle) and 16.8~GeV (right). The estimates are based on Eq. \ref{eq:dt_standard}. UrQMD calculations are shown as full lines (dashed lines) for the ratio $K^{*0}/K^+$ ($K^{*0}/K^-$), open circles (filled squares) show experimental $K^{*0}/K^+$ ($K^{*0}/K^-$) ratios converted to $\gamma\Delta t$ \cite{Kozlowski:2024cjw} using Eq.~\ref{eq:dt_standard}.

One generally observes that the estimated lifetime ($\gamma\Delta t$) of the hadronic stage increases with system size, both in the model and in the experimental data. However, it seems that starting from $\sqrt{s_\mathrm{NN}} = $ 11.9~GeV, the lifetime for heavier systems ($N_\mathrm{part}\geq 60$) in the data is substantially larger than in the model. If this is confirmed, it may hint to a softer expansion that might indicate a critical point/first order phase transition. 

One should note that the estimates using the full rate equation leads to substantially longer lifetimes, which can reach up to approx 10 fm/c, significantly larger than with the simplified method used by the experiment.

\section{Conclusion}

We have studied the suppression of $K^*(892)$ resonances relative to ground-state Kaons in Ar+Sc collisions at SPS energies and utilized them as a tool to estimate the lifetime of the fireball in intermediate mass systems. For the analysis, we employed the Ultra-relativistic Quantum Molecular Dynamics model to calculate the hadronic evolution in Ar+Sc collisions to obtain the yields and spectra of observable Kaon resonances. The yields and spectra of reconstructable $K^*(892)$ resonances are generally in good agreement with the data measured by the NA61/SHINE collaboration. Finally, we compared to the lifetime estimates reported by the experimental collaboration and found good but not perfect agreement with the data, especially in very central collisions, the experimental lifetimes seem to be substantially longer.

\section*{Acknowledgments}
T.R. acknowledges support from The Branco Weiss Fellowship - Society in Science, administered by the ETH Z\"urich.
The computational resources for this project were provided by the Center for Scientific Computing of the GU Frankfurt and the Goethe HLR.

\clearpage
\onecolumngrid
\appendix
\section{Numerical data}

\subsection{Numerical data used in Fig.~\ref{fig:plot_y_dNdy_all} for $\sqrt{s_{NN}}=8.8, 11.9, 16.8$~GeV}

\scriptsize
\setlength{\LTleft}{0pt}
\setlength{\LTright}{0pt}

\begin{longtable}{c c c c c c c c c}
\caption{Numerical values used for the UrQMD rapidity distributions in Ar+Sc collisions at $\sqrt{s_{NN}}=8.8$~GeV. The table lists the rapidity bin center $y$, the yields, and the corresponding statistical uncertainties for $K^+$, $K^-$, $K^{*0}$, and $\bar{K}^{*0}$.}
\label{tab:dndy_ecm8p8}\\
\toprule
$y$ & $K^+$ & $\Delta K^+$ & $K^-$ & $\Delta K^-$ & $K^{*0}$ & $\Delta K^{*0}$ & $\bar{K}^{*0}$ & $\Delta \bar{K}^{*0}$ \\
\midrule
\endfirsthead

\multicolumn{9}{c}{\tablename\ \thetable\ -- continued from previous page} \\
\toprule
$y$ & $K^+$ & $\Delta K^+$ & $K^-$ & $\Delta K^-$ & $K^{*0}$ & $\Delta K^{*0}$ & $\bar{K}^{*0}$ & $\Delta \bar{K}^{*0}$ \\
\midrule
\endhead

\midrule
\multicolumn{9}{r}{continued on next page} \\
\endfoot

\bottomrule
\endlastfoot

-3.9500 & 0.0000e+00 & 0.0000e+00 & 0.0000e+00 & 0.0000e+00 & 0.0000e+00 & 0.0000e+00 & 0.0000e+00 & 0.0000e+00 \\
-3.8500 & 0.0000e+00 & 0.0000e+00 & 0.0000e+00 & 0.0000e+00 & 0.0000e+00 & 0.0000e+00 & 0.0000e+00 & 0.0000e+00 \\
-3.7500 & 0.0000e+00 & 0.0000e+00 & 0.0000e+00 & 0.0000e+00 & 0.0000e+00 & 0.0000e+00 & 0.0000e+00 & 0.0000e+00 \\
-3.6500 & 0.0000e+00 & 0.0000e+00 & 0.0000e+00 & 0.0000e+00 & 0.0000e+00 & 0.0000e+00 & 0.0000e+00 & 0.0000e+00 \\
-3.5500 & 0.0000e+00 & 0.0000e+00 & 0.0000e+00 & 0.0000e+00 & 0.0000e+00 & 0.0000e+00 & 0.0000e+00 & 0.0000e+00 \\
-3.4500 & 0.0000e+00 & 0.0000e+00 & 0.0000e+00 & 0.0000e+00 & 0.0000e+00 & 0.0000e+00 & 0.0000e+00 & 0.0000e+00 \\
-3.3500 & 2.0000e-05 & 2.0000e-05 & 0.0000e+00 & 0.0000e+00 & 0.0000e+00 & 0.0000e+00 & 0.0000e+00 & 0.0000e+00 \\
-3.2500 & 6.0000e-05 & 3.4641e-05 & 0.0000e+00 & 0.0000e+00 & 0.0000e+00 & 0.0000e+00 & 0.0000e+00 & 0.0000e+00 \\
-3.1500 & 2.2000e-04 & 6.6332e-05 & 4.0000e-05 & 2.8284e-05 & 0.0000e+00 & 0.0000e+00 & 0.0000e+00 & 0.0000e+00 \\
-3.0500 & 3.4000e-04 & 8.2462e-05 & 1.2000e-04 & 4.8990e-05 & 0.0000e+00 & 0.0000e+00 & 0.0000e+00 & 0.0000e+00 \\
-2.9500 & 1.4000e-03 & 1.6733e-04 & 3.4000e-04 & 8.2462e-05 & 2.0000e-05 & 2.0000e-05 & 0.0000e+00 & 0.0000e+00 \\
-2.8500 & 3.2400e-03 & 2.5456e-04 & 1.0600e-03 & 1.4560e-04 & 2.0000e-05 & 2.0000e-05 & 2.0000e-05 & 2.0000e-05 \\
-2.7500 & 7.6000e-03 & 3.8987e-04 & 1.3600e-03 & 1.6492e-04 & 8.0000e-05 & 4.0000e-05 & 2.0000e-05 & 2.0000e-05 \\
-2.6500 & 1.5860e-02 & 5.6321e-04 & 3.3800e-03 & 2.6000e-04 & 3.4000e-04 & 8.2462e-05 & 1.2000e-04 & 4.8990e-05 \\
-2.5500 & 3.1520e-02 & 7.9398e-04 & 6.1400e-03 & 3.5043e-04 & 8.4000e-04 & 1.2961e-04 & 1.4000e-04 & 5.2915e-05 \\
-2.4500 & 5.6400e-02 & 1.0621e-03 & 1.1200e-02 & 4.7329e-04 & 1.7400e-03 & 1.8655e-04 & 3.6000e-04 & 8.4853e-05 \\
-2.3500 & 9.3180e-02 & 1.3651e-03 & 1.9000e-02 & 6.1644e-04 & 3.7600e-03 & 2.7423e-04 & 4.6000e-04 & 9.5917e-05 \\
-2.2500 & 1.4798e-01 & 1.7203e-03 & 3.1340e-02 & 7.9171e-04 & 9.1400e-03 & 4.2755e-04 & 1.4000e-03 & 1.6733e-04 \\
-2.1500 & 2.1904e-01 & 2.0930e-03 & 4.5880e-02 & 9.5791e-04 & 1.8400e-02 & 6.0663e-04 & 3.0200e-03 & 2.4576e-04 \\
-2.0500 & 3.0242e-01 & 2.4593e-03 & 6.9480e-02 & 1.1788e-03 & 3.4000e-02 & 8.2462e-04 & 5.2600e-03 & 3.2435e-04 \\
-1.9500 & 4.1274e-01 & 2.8731e-03 & 9.9720e-02 & 1.4122e-03 & 5.9200e-02 & 1.0881e-03 & 9.0400e-03 & 4.2521e-04 \\
-1.8500 & 5.3470e-01 & 3.2702e-03 & 1.3938e-01 & 1.6696e-03 & 9.0480e-02 & 1.3452e-03 & 1.4080e-02 & 5.3066e-04 \\
-1.7500 & 6.7184e-01 & 3.6656e-03 & 1.8788e-01 & 1.9385e-03 & 1.2994e-01 & 1.6121e-03 & 2.2400e-02 & 6.6933e-04 \\
-1.6500 & 8.0912e-01 & 4.0227e-03 & 2.4392e-01 & 2.2087e-03 & 1.7434e-01 & 1.8673e-03 & 3.0760e-02 & 7.8435e-04 \\
-1.5500 & 9.6956e-01 & 4.4035e-03 & 3.1148e-01 & 2.4959e-03 & 2.2686e-01 & 2.1301e-03 & 4.4100e-02 & 9.3915e-04 \\
-1.4500 & 1.1326e+00 & 4.7594e-03 & 3.9094e-01 & 2.7962e-03 & 2.7194e-01 & 2.3321e-03 & 5.9340e-02 & 1.0894e-03 \\
-1.3500 & 1.2803e+00 & 5.0602e-03 & 4.8144e-01 & 3.1030e-03 & 3.2060e-01 & 2.5322e-03 & 7.5740e-02 & 1.2308e-03 \\
-1.2500 & 1.4266e+00 & 5.3416e-03 & 5.7620e-01 & 3.3947e-03 & 3.6522e-01 & 2.7027e-03 & 9.5640e-02 & 1.3830e-03 \\
-1.1500 & 1.5736e+00 & 5.6100e-03 & 6.7852e-01 & 3.6838e-03 & 4.1222e-01 & 2.8713e-03 & 1.1514e-01 & 1.5175e-03 \\
-1.0500 & 1.6921e+00 & 5.8174e-03 & 7.7564e-01 & 3.9386e-03 & 4.5210e-01 & 3.0070e-03 & 1.3642e-01 & 1.6518e-03 \\
-0.9500 & 1.8033e+00 & 6.0056e-03 & 8.8452e-01 & 4.2060e-03 & 4.8724e-01 & 3.1217e-03 & 1.5866e-01 & 1.7813e-03 \\
-0.8500 & 1.9140e+00 & 6.1871e-03 & 9.9054e-01 & 4.4509e-03 & 5.2340e-01 & 3.2354e-03 & 1.8040e-01 & 1.8995e-03 \\
-0.7500 & 2.0240e+00 & 6.3624e-03 & 1.0905e+00 & 4.6700e-03 & 5.5728e-01 & 3.3385e-03 & 2.0710e-01 & 2.0352e-03 \\
-0.6500 & 2.1077e+00 & 6.4926e-03 & 1.1864e+00 & 4.8711e-03 & 5.8458e-01 & 3.4193e-03 & 2.2776e-01 & 2.1343e-03 \\
-0.5500 & 2.1773e+00 & 6.5989e-03 & 1.2653e+00 & 5.0306e-03 & 6.0518e-01 & 3.4790e-03 & 2.4218e-01 & 2.2008e-03 \\
-0.4500 & 2.2533e+00 & 6.7131e-03 & 1.3500e+00 & 5.1962e-03 & 6.2178e-01 & 3.5264e-03 & 2.6420e-01 & 2.2987e-03 \\
-0.3500 & 2.3026e+00 & 6.7862e-03 & 1.4150e+00 & 5.3197e-03 & 6.4172e-01 & 3.5825e-03 & 2.7604e-01 & 2.3496e-03 \\
-0.2500 & 2.3463e+00 & 6.8503e-03 & 1.4634e+00 & 5.4100e-03 & 6.5228e-01 & 3.6119e-03 & 2.8656e-01 & 2.3940e-03 \\
-0.1500 & 2.3691e+00 & 6.8834e-03 & 1.4893e+00 & 5.4576e-03 & 6.5916e-01 & 3.6309e-03 & 2.9460e-01 & 2.4273e-03 \\
-0.0500 & 2.3704e+00 & 6.8853e-03 & 1.5041e+00 & 5.4846e-03 & 6.6186e-01 & 3.6383e-03 & 2.9992e-01 & 2.4492e-03 \\
0.0500 & 2.3591e+00 & 6.8689e-03 & 1.5090e+00 & 5.4936e-03 & 6.6304e-01 & 3.6415e-03 & 2.9736e-01 & 2.4387e-03 \\
0.1500 & 2.3532e+00 & 6.8604e-03 & 1.4905e+00 & 5.4598e-03 & 6.5226e-01 & 3.6118e-03 & 2.9212e-01 & 2.4171e-03 \\
0.2500 & 2.3081e+00 & 6.7943e-03 & 1.4468e+00 & 5.3793e-03 & 6.4662e-01 & 3.5962e-03 & 2.8446e-01 & 2.3852e-03 \\
0.3500 & 2.2620e+00 & 6.7261e-03 & 1.3981e+00 & 5.2880e-03 & 6.3324e-01 & 3.5588e-03 & 2.7446e-01 & 2.3429e-03 \\
0.4500 & 2.2068e+00 & 6.6436e-03 & 1.3434e+00 & 5.1834e-03 & 6.1362e-01 & 3.5032e-03 & 2.6112e-01 & 2.2853e-03 \\
0.5500 & 2.1414e+00 & 6.5443e-03 & 1.2446e+00 & 4.9892e-03 & 5.8660e-01 & 3.4252e-03 & 2.4302e-01 & 2.2046e-03 \\
0.6500 & 2.0507e+00 & 6.4042e-03 & 1.1626e+00 & 4.8220e-03 & 5.6202e-01 & 3.3527e-03 & 2.2336e-01 & 2.1136e-03 \\
0.7500 & 1.9321e+00 & 6.2162e-03 & 1.0707e+00 & 4.6275e-03 & 5.3372e-01 & 3.2672e-03 & 2.0398e-01 & 2.0198e-03 \\
0.8500 & 1.8448e+00 & 6.0743e-03 & 9.5940e-01 & 4.3804e-03 & 5.0304e-01 & 3.1719e-03 & 1.7442e-01 & 1.8677e-03 \\
0.9500 & 1.7272e+00 & 5.8774e-03 & 8.5796e-01 & 4.1424e-03 & 4.7540e-01 & 3.0835e-03 & 1.5866e-01 & 1.7813e-03 \\
1.0500 & 1.6053e+00 & 5.6661e-03 & 7.4488e-01 & 3.8597e-03 & 4.3240e-01 & 2.9407e-03 & 1.3090e-01 & 1.6180e-03 \\
1.1500 & 1.4775e+00 & 5.4360e-03 & 6.4012e-01 & 3.5780e-03 & 3.9320e-01 & 2.8043e-03 & 1.0770e-01 & 1.4677e-03 \\
1.2500 & 1.3438e+00 & 5.1842e-03 & 5.4282e-01 & 3.2949e-03 & 3.4882e-01 & 2.6413e-03 & 8.8280e-02 & 1.3288e-03 \\
1.3500 & 1.1927e+00 & 4.8841e-03 & 4.5074e-01 & 3.0025e-03 & 3.0578e-01 & 2.4730e-03 & 7.2000e-02 & 1.2000e-03 \\
1.4500 & 1.0554e+00 & 4.5943e-03 & 3.6532e-01 & 2.7030e-03 & 2.5938e-01 & 2.2776e-03 & 5.5260e-02 & 1.0513e-03 \\
1.5500 & 9.0978e-01 & 4.2656e-03 & 2.8840e-01 & 2.4017e-03 & 2.0814e-01 & 2.0403e-03 & 4.1300e-02 & 9.0885e-04 \\
1.6500 & 7.5502e-01 & 3.8859e-03 & 2.2072e-01 & 2.1010e-03 & 1.6384e-01 & 1.8102e-03 & 2.9760e-02 & 7.7149e-04 \\
1.7500 & 6.1782e-01 & 3.5152e-03 & 1.6840e-01 & 1.8352e-03 & 1.2288e-01 & 1.5677e-03 & 1.8520e-02 & 6.0860e-04 \\
1.8500 & 4.9104e-01 & 3.1338e-03 & 1.2126e-01 & 1.5573e-03 & 8.3780e-02 & 1.2944e-03 & 1.3080e-02 & 5.1147e-04 \\
1.9500 & 3.7644e-01 & 2.7439e-03 & 8.8160e-02 & 1.3279e-03 & 5.2540e-02 & 1.0251e-03 & 8.4200e-03 & 4.1037e-04 \\
2.0500 & 2.7658e-01 & 2.3519e-03 & 6.0200e-02 & 1.0973e-03 & 3.1000e-02 & 7.8740e-04 & 4.3400e-03 & 2.9462e-04 \\
2.1500 & 1.9544e-01 & 1.9771e-03 & 4.0880e-02 & 9.0421e-04 & 1.6200e-02 & 5.6921e-04 & 2.3600e-03 & 2.1726e-04 \\
2.2500 & 1.3166e-01 & 1.6227e-03 & 2.6140e-02 & 7.2305e-04 & 7.9600e-03 & 3.9900e-04 & 1.0800e-03 & 1.4697e-04 \\
2.3500 & 8.4380e-02 & 1.2991e-03 & 1.5780e-02 & 5.6178e-04 & 3.3000e-03 & 2.5690e-04 & 4.4000e-04 & 9.3808e-05 \\
2.4500 & 5.0600e-02 & 1.0060e-03 & 8.5200e-03 & 4.1280e-04 & 1.5800e-03 & 1.7776e-04 & 1.8000e-04 & 6.0000e-05 \\
2.5500 & 2.7800e-02 & 7.4565e-04 & 4.8200e-03 & 3.1048e-04 & 7.8000e-04 & 1.2490e-04 & 1.4000e-04 & 5.2915e-05 \\
2.6500 & 1.4760e-02 & 5.4332e-04 & 2.4600e-03 & 2.2181e-04 & 1.4000e-04 & 5.2915e-05 & 2.0000e-05 & 2.0000e-05 \\
2.7500 & 5.9000e-03 & 3.4351e-04 & 1.2800e-03 & 1.6000e-04 & 4.0000e-05 & 2.8284e-05 & 0.0000e+00 & 0.0000e+00 \\
2.8500 & 3.1000e-03 & 2.4900e-04 & 6.0000e-04 & 1.0954e-04 & 0.0000e+00 & 0.0000e+00 & 0.0000e+00 & 0.0000e+00 \\
2.9500 & 1.2400e-03 & 1.5748e-04 & 2.8000e-04 & 7.4833e-05 & 0.0000e+00 & 0.0000e+00 & 2.0000e-05 & 2.0000e-05 \\
3.0500 & 4.0000e-04 & 8.9443e-05 & 1.2000e-04 & 4.8990e-05 & 0.0000e+00 & 0.0000e+00 & 0.0000e+00 & 0.0000e+00 \\
3.1500 & 1.6000e-04 & 5.6569e-05 & 2.0000e-05 & 2.0000e-05 & 0.0000e+00 & 0.0000e+00 & 0.0000e+00 & 0.0000e+00 \\
3.2500 & 4.0000e-05 & 2.8284e-05 & 0.0000e+00 & 0.0000e+00 & 0.0000e+00 & 0.0000e+00 & 0.0000e+00 & 0.0000e+00 \\
3.3500 & 2.0000e-05 & 2.0000e-05 & 0.0000e+00 & 0.0000e+00 & 0.0000e+00 & 0.0000e+00 & 0.0000e+00 & 0.0000e+00 \\
3.4500 & 0.0000e+00 & 0.0000e+00 & 0.0000e+00 & 0.0000e+00 & 0.0000e+00 & 0.0000e+00 & 0.0000e+00 & 0.0000e+00 \\
3.5500 & 0.0000e+00 & 0.0000e+00 & 0.0000e+00 & 0.0000e+00 & 0.0000e+00 & 0.0000e+00 & 0.0000e+00 & 0.0000e+00 \\
3.6500 & 0.0000e+00 & 0.0000e+00 & 0.0000e+00 & 0.0000e+00 & 0.0000e+00 & 0.0000e+00 & 0.0000e+00 & 0.0000e+00 \\
3.7500 & 0.0000e+00 & 0.0000e+00 & 0.0000e+00 & 0.0000e+00 & 0.0000e+00 & 0.0000e+00 & 0.0000e+00 & 0.0000e+00 \\
3.8500 & 0.0000e+00 & 0.0000e+00 & 0.0000e+00 & 0.0000e+00 & 0.0000e+00 & 0.0000e+00 & 0.0000e+00 & 0.0000e+00 \\
3.9500 & 0.0000e+00 & 0.0000e+00 & 0.0000e+00 & 0.0000e+00 & 0.0000e+00 & 0.0000e+00 & 0.0000e+00 & 0.0000e+00 \\
\end{longtable}

\scriptsize
\setlength{\LTleft}{0pt}
\setlength{\LTright}{0pt}

\begin{longtable}{c c c c c c c c c}
\caption{Numerical values used for the UrQMD rapidity distributions in Ar+Sc collisions at $\sqrt{s_{NN}}=11.9$~GeV. The table lists the rapidity bin center $y$, the yields, and the corresponding statistical uncertainties for $K^+$, $K^-$, $K^{*0}$, and $\bar{K}^{*0}$.}
\label{tab:dndy_ecm11p9}\\
\toprule
$y$ & $K^+$ & $\Delta K^+$ & $K^-$ & $\Delta K^-$ & $K^{*0}$ & $\Delta K^{*0}$ & $\bar{K}^{*0}$ & $\Delta \bar{K}^{*0}$ \\
\midrule
\endfirsthead

\multicolumn{9}{c}{\tablename\ \thetable\ -- continued from previous page} \\
\toprule
$y$ & $K^+$ & $\Delta K^+$ & $K^-$ & $\Delta K^-$ & $K^{*0}$ & $\Delta K^{*0}$ & $\bar{K}^{*0}$ & $\Delta \bar{K}^{*0}$ \\
\midrule
\endhead

\midrule
\multicolumn{9}{r}{continued on next page} \\
\endfoot

\bottomrule
\endlastfoot

-3.9500 & 0.0000e+00 & 0.0000e+00 & 0.0000e+00 & 0.0000e+00 & 0.0000e+00 & 0.0000e+00 & 0.0000e+00 & 0.0000e+00 \\
-3.8500 & 0.0000e+00 & 0.0000e+00 & 0.0000e+00 & 0.0000e+00 & 0.0000e+00 & 0.0000e+00 & 0.0000e+00 & 0.0000e+00 \\
-3.7500 & 0.0000e+00 & 0.0000e+00 & 0.0000e+00 & 0.0000e+00 & 0.0000e+00 & 0.0000e+00 & 0.0000e+00 & 0.0000e+00 \\
-3.6500 & 2.0000e-05 & 2.0000e-05 & 0.0000e+00 & 0.0000e+00 & 0.0000e+00 & 0.0000e+00 & 0.0000e+00 & 0.0000e+00 \\
-3.5500 & 4.0000e-05 & 2.8284e-05 & 0.0000e+00 & 0.0000e+00 & 0.0000e+00 & 0.0000e+00 & 0.0000e+00 & 0.0000e+00 \\
-3.4500 & 1.2000e-04 & 4.8990e-05 & 4.0000e-05 & 2.8284e-05 & 0.0000e+00 & 0.0000e+00 & 0.0000e+00 & 0.0000e+00 \\
-3.3500 & 4.8000e-04 & 9.7980e-05 & 1.6000e-04 & 5.6569e-05 & 0.0000e+00 & 0.0000e+00 & 0.0000e+00 & 0.0000e+00 \\
-3.2500 & 1.2600e-03 & 1.5875e-04 & 4.0000e-04 & 8.9443e-05 & 2.0000e-05 & 2.0000e-05 & 0.0000e+00 & 0.0000e+00 \\
-3.1500 & 3.2400e-03 & 2.5456e-04 & 7.2000e-04 & 1.2000e-04 & 0.0000e+00 & 0.0000e+00 & 0.0000e+00 & 0.0000e+00 \\
-3.0500 & 8.2000e-03 & 4.0497e-04 & 1.4800e-03 & 1.7205e-04 & 6.0000e-05 & 3.4641e-05 & 0.0000e+00 & 0.0000e+00 \\
-2.9500 & 1.6440e-02 & 5.7341e-04 & 3.2600e-03 & 2.5534e-04 & 4.8000e-04 & 9.7980e-05 & 4.0000e-05 & 2.8284e-05 \\
-2.8500 & 3.1320e-02 & 7.9145e-04 & 6.4200e-03 & 3.5833e-04 & 7.2000e-04 & 1.2000e-04 & 1.6000e-04 & 5.6569e-05 \\
-2.7500 & 5.7880e-02 & 1.0759e-03 & 1.1940e-02 & 4.8867e-04 & 2.0200e-03 & 2.0100e-04 & 3.2000e-04 & 8.0000e-05 \\
-2.6500 & 9.4400e-02 & 1.3740e-03 & 2.0200e-02 & 6.3561e-04 & 4.6000e-03 & 3.0332e-04 & 8.2000e-04 & 1.2806e-04 \\
-2.5500 & 1.5158e-01 & 1.7411e-03 & 3.2400e-02 & 8.0498e-04 & 9.5200e-03 & 4.3635e-04 & 1.5000e-03 & 1.7321e-04 \\
-2.4500 & 2.1850e-01 & 2.0905e-03 & 5.0400e-02 & 1.0040e-03 & 1.8980e-02 & 6.1612e-04 & 3.2200e-03 & 2.5377e-04 \\
-2.3500 & 3.1324e-01 & 2.5030e-03 & 7.6020e-02 & 1.2330e-03 & 3.6820e-02 & 8.5814e-04 & 5.3400e-03 & 3.2680e-04 \\
-2.2500 & 4.1912e-01 & 2.8952e-03 & 1.0902e-01 & 1.4766e-03 & 6.1820e-02 & 1.1119e-03 & 9.9800e-03 & 4.4677e-04 \\
-2.1500 & 5.5052e-01 & 3.3182e-03 & 1.4706e-01 & 1.7150e-03 & 9.6260e-02 & 1.3875e-03 & 1.5600e-02 & 5.5857e-04 \\
-2.0500 & 6.9698e-01 & 3.7336e-03 & 1.9796e-01 & 1.9898e-03 & 1.3822e-01 & 1.6626e-03 & 2.6240e-02 & 7.2443e-04 \\
-1.9500 & 8.4802e-01 & 4.1183e-03 & 2.6662e-01 & 2.3092e-03 & 1.8560e-01 & 1.9267e-03 & 3.6380e-02 & 8.5299e-04 \\
-1.8500 & 1.0135e+00 & 4.5022e-03 & 3.4576e-01 & 2.6297e-03 & 2.3972e-01 & 2.1896e-03 & 5.2080e-02 & 1.0206e-03 \\
-1.7500 & 1.1749e+00 & 4.8476e-03 & 4.4138e-01 & 2.9711e-03 & 2.9414e-01 & 2.4254e-03 & 7.0540e-02 & 1.1878e-03 \\
-1.6500 & 1.3430e+00 & 5.1826e-03 & 5.3512e-01 & 3.2715e-03 & 3.4342e-01 & 2.6208e-03 & 9.1660e-02 & 1.3540e-03 \\
-1.5500 & 1.4995e+00 & 5.4764e-03 & 6.4718e-01 & 3.5977e-03 & 3.9566e-01 & 2.8130e-03 & 1.1598e-01 & 1.5230e-03 \\
-1.4500 & 1.6584e+00 & 5.7592e-03 & 7.5498e-01 & 3.8858e-03 & 4.4648e-01 & 2.9882e-03 & 1.4166e-01 & 1.6832e-03 \\
-1.3500 & 1.7930e+00 & 5.9883e-03 & 8.8076e-01 & 4.1970e-03 & 4.8718e-01 & 3.1215e-03 & 1.6884e-01 & 1.8376e-03 \\
-1.2500 & 1.9319e+00 & 6.2160e-03 & 1.0056e+00 & 4.4846e-03 & 5.3476e-01 & 3.2704e-03 & 1.9668e-01 & 1.9833e-03 \\
-1.1500 & 2.0447e+00 & 6.3948e-03 & 1.1303e+00 & 4.7545e-03 & 5.7066e-01 & 3.3783e-03 & 2.2736e-01 & 2.1324e-03 \\
-1.0500 & 2.1664e+00 & 6.5825e-03 & 1.2638e+00 & 5.0276e-03 & 6.1232e-01 & 3.4995e-03 & 2.6000e-01 & 2.2804e-03 \\
-0.9500 & 2.2847e+00 & 6.7597e-03 & 1.3743e+00 & 5.2427e-03 & 6.5218e-01 & 3.6116e-03 & 2.8584e-01 & 2.3910e-03 \\
-0.8500 & 2.3858e+00 & 6.9077e-03 & 1.5005e+00 & 5.4781e-03 & 6.8448e-01 & 3.6999e-03 & 3.1314e-01 & 2.5026e-03 \\
-0.7500 & 2.4694e+00 & 7.0277e-03 & 1.6060e+00 & 5.6675e-03 & 7.0296e-01 & 3.7496e-03 & 3.4444e-01 & 2.6247e-03 \\
-0.6500 & 2.5533e+00 & 7.1460e-03 & 1.7018e+00 & 5.8341e-03 & 7.2956e-01 & 3.8198e-03 & 3.7196e-01 & 2.7275e-03 \\
-0.5500 & 2.6187e+00 & 7.2370e-03 & 1.7659e+00 & 5.9429e-03 & 7.4330e-01 & 3.8556e-03 & 3.9412e-01 & 2.8076e-03 \\
-0.4500 & 2.6520e+00 & 7.2828e-03 & 1.8566e+00 & 6.0936e-03 & 7.6980e-01 & 3.9238e-03 & 4.1340e-01 & 2.8754e-03 \\
-0.3500 & 2.7134e+00 & 7.3667e-03 & 1.9063e+00 & 6.1746e-03 & 7.8324e-01 & 3.9579e-03 & 4.3124e-01 & 2.9368e-03 \\
-0.2500 & 2.7461e+00 & 7.4110e-03 & 1.9515e+00 & 6.2474e-03 & 7.9202e-01 & 3.9800e-03 & 4.3578e-01 & 2.9522e-03 \\
-0.1500 & 2.7757e+00 & 7.4508e-03 & 1.9781e+00 & 6.2898e-03 & 7.9756e-01 & 3.9939e-03 & 4.4882e-01 & 2.9961e-03 \\
-0.0500 & 2.7812e+00 & 7.4581e-03 & 1.9797e+00 & 6.2923e-03 & 8.0246e-01 & 4.0061e-03 & 4.4714e-01 & 2.9905e-03 \\
0.0500 & 2.7865e+00 & 7.4652e-03 & 1.9894e+00 & 6.3078e-03 & 8.0236e-01 & 4.0059e-03 & 4.4344e-01 & 2.9781e-03 \\
0.1500 & 2.7476e+00 & 7.4129e-03 & 1.9588e+00 & 6.2591e-03 & 8.0228e-01 & 4.0057e-03 & 4.4010e-01 & 2.9668e-03 \\
0.2500 & 2.7268e+00 & 7.3848e-03 & 1.9303e+00 & 6.2133e-03 & 7.8118e-01 & 3.9527e-03 & 4.3048e-01 & 2.9342e-03 \\
0.3500 & 2.6840e+00 & 7.3267e-03 & 1.8842e+00 & 6.1387e-03 & 7.7448e-01 & 3.9357e-03 & 4.2508e-01 & 2.9158e-03 \\
0.4500 & 2.6270e+00 & 7.2484e-03 & 1.8295e+00 & 6.0490e-03 & 7.5992e-01 & 3.8985e-03 & 4.0608e-01 & 2.8498e-03 \\
0.5500 & 2.5583e+00 & 7.1530e-03 & 1.7549e+00 & 5.9244e-03 & 7.3332e-01 & 3.8297e-03 & 3.8822e-01 & 2.7865e-03 \\
0.6500 & 2.4790e+00 & 7.0414e-03 & 1.6697e+00 & 5.7788e-03 & 7.1956e-01 & 3.7936e-03 & 3.6926e-01 & 2.7176e-03 \\
0.7500 & 2.3866e+00 & 6.9089e-03 & 1.5710e+00 & 5.6053e-03 & 6.8998e-01 & 3.7148e-03 & 3.4024e-01 & 2.6086e-03 \\
0.8500 & 2.3031e+00 & 6.7869e-03 & 1.4601e+00 & 5.4039e-03 & 6.6376e-01 & 3.6435e-03 & 3.1422e-01 & 2.5069e-03 \\
0.9500 & 2.2010e+00 & 6.6348e-03 & 1.3473e+00 & 5.1910e-03 & 6.2160e-01 & 3.5259e-03 & 2.8488e-01 & 2.3870e-03 \\
1.0500 & 2.0857e+00 & 6.4586e-03 & 1.2285e+00 & 4.9569e-03 & 5.8774e-01 & 3.4285e-03 & 2.5428e-01 & 2.2551e-03 \\
1.1500 & 1.9599e+00 & 6.2608e-03 & 1.0990e+00 & 4.6884e-03 & 5.5274e-01 & 3.3249e-03 & 2.2472e-01 & 2.1200e-03 \\
1.2500 & 1.8239e+00 & 6.0397e-03 & 9.6626e-01 & 4.3960e-03 & 5.1156e-01 & 3.1986e-03 & 1.9326e-01 & 1.9660e-03 \\
1.3500 & 1.7070e+00 & 5.8429e-03 & 8.4336e-01 & 4.1070e-03 & 4.6654e-01 & 3.0546e-03 & 1.6328e-01 & 1.8071e-03 \\
1.4500 & 1.5507e+00 & 5.5691e-03 & 7.2498e-01 & 3.8078e-03 & 4.2378e-01 & 2.9113e-03 & 1.3604e-01 & 1.6495e-03 \\
1.5500 & 1.4166e+00 & 5.3227e-03 & 6.0634e-01 & 3.4824e-03 & 3.8138e-01 & 2.7618e-03 & 1.1250e-01 & 1.5000e-03 \\
1.6500 & 1.2558e+00 & 5.0115e-03 & 5.0462e-01 & 3.1769e-03 & 3.2280e-01 & 2.5409e-03 & 8.7280e-02 & 1.3212e-03 \\
1.7500 & 1.0963e+00 & 4.6825e-03 & 4.0302e-01 & 2.8391e-03 & 2.7378e-01 & 2.3400e-03 & 6.7620e-02 & 1.1629e-03 \\
1.8500 & 9.4410e-01 & 4.3453e-03 & 3.2060e-01 & 2.5322e-03 & 2.2748e-01 & 2.1330e-03 & 4.9840e-02 & 9.9840e-04 \\
1.9500 & 7.8088e-01 & 3.9519e-03 & 2.4538e-01 & 2.2153e-03 & 1.7266e-01 & 1.8583e-03 & 3.4840e-02 & 8.3475e-04 \\
2.0500 & 6.3838e-01 & 3.5732e-03 & 1.8074e-01 & 1.9013e-03 & 1.2628e-01 & 1.5892e-03 & 2.3220e-02 & 6.8147e-04 \\
2.1500 & 5.0732e-01 & 3.1853e-03 & 1.3552e-01 & 1.6463e-03 & 8.7600e-02 & 1.3236e-03 & 1.5340e-02 & 5.5390e-04 \\
2.2500 & 3.8236e-01 & 2.7654e-03 & 9.4160e-02 & 1.3723e-03 & 5.5680e-02 & 1.0553e-03 & 9.9000e-03 & 4.4497e-04 \\
2.3500 & 2.8006e-01 & 2.3667e-03 & 6.5700e-02 & 1.1463e-03 & 3.1180e-02 & 7.8968e-04 & 5.4800e-03 & 3.3106e-04 \\
2.4500 & 1.9858e-01 & 1.9929e-03 & 4.4580e-02 & 9.4425e-04 & 1.7180e-02 & 5.8617e-04 & 2.7200e-03 & 2.3324e-04 \\
2.5500 & 1.3314e-01 & 1.6318e-03 & 2.6640e-02 & 7.2993e-04 & 8.2800e-03 & 4.0694e-04 & 1.5400e-03 & 1.7550e-04 \\
2.6500 & 8.2340e-02 & 1.2833e-03 & 1.6520e-02 & 5.7480e-04 & 3.9600e-03 & 2.8142e-04 & 3.4000e-04 & 8.2462e-05 \\
2.7500 & 4.9220e-02 & 9.9217e-04 & 1.0020e-02 & 4.4766e-04 & 1.9600e-03 & 1.9799e-04 & 3.4000e-04 & 8.2462e-05 \\
2.8500 & 2.7880e-02 & 7.4673e-04 & 4.4800e-03 & 2.9933e-04 & 5.0000e-04 & 1.0000e-04 & 1.0000e-04 & 4.4721e-05 \\
2.9500 & 1.2980e-02 & 5.0951e-04 & 2.6000e-03 & 2.2804e-04 & 1.6000e-04 & 5.6569e-05 & 4.0000e-05 & 2.8284e-05 \\
3.0500 & 6.6000e-03 & 3.6332e-04 & 1.2800e-03 & 1.6000e-04 & 2.0000e-05 & 2.0000e-05 & 0.0000e+00 & 0.0000e+00 \\
3.1500 & 2.7600e-03 & 2.3495e-04 & 6.8000e-04 & 1.1662e-04 & 2.0000e-05 & 2.0000e-05 & 0.0000e+00 & 0.0000e+00 \\
3.2500 & 1.0800e-03 & 1.4697e-04 & 2.6000e-04 & 7.2111e-05 & 2.0000e-05 & 2.0000e-05 & 0.0000e+00 & 0.0000e+00 \\
3.3500 & 3.4000e-04 & 8.2462e-05 & 1.8000e-04 & 6.0000e-05 & 0.0000e+00 & 0.0000e+00 & 0.0000e+00 & 0.0000e+00 \\
3.4500 & 6.0000e-05 & 3.4641e-05 & 0.0000e+00 & 0.0000e+00 & 0.0000e+00 & 0.0000e+00 & 0.0000e+00 & 0.0000e+00 \\
3.5500 & 2.0000e-05 & 2.0000e-05 & 4.0000e-05 & 2.8284e-05 & 0.0000e+00 & 0.0000e+00 & 0.0000e+00 & 0.0000e+00 \\
3.6500 & 4.0000e-05 & 2.8284e-05 & 0.0000e+00 & 0.0000e+00 & 0.0000e+00 & 0.0000e+00 & 0.0000e+00 & 0.0000e+00 \\
3.7500 & 0.0000e+00 & 0.0000e+00 & 0.0000e+00 & 0.0000e+00 & 0.0000e+00 & 0.0000e+00 & 0.0000e+00 & 0.0000e+00 \\
3.8500 & 0.0000e+00 & 0.0000e+00 & 0.0000e+00 & 0.0000e+00 & 0.0000e+00 & 0.0000e+00 & 0.0000e+00 & 0.0000e+00 \\
3.9500 & 0.0000e+00 & 0.0000e+00 & 0.0000e+00 & 0.0000e+00 & 0.0000e+00 & 0.0000e+00 & 0.0000e+00 & 0.0000e+00 \\
\end{longtable}

\scriptsize
\setlength{\LTleft}{0pt}
\setlength{\LTright}{0pt}

\begin{longtable}{c c c c c c c c c}
\caption{Numerical values used for the UrQMD rapidity distributions in Ar+Sc collisions at $\sqrt{s_{NN}}=16.8$~GeV. The table lists the rapidity bin center $y$, the yields, and the corresponding statistical uncertainties for $K^+$, $K^-$, $K^{*0}$, and $\bar{K}^{*0}$.}
\label{tab:dndy_ecm16p8}\\
\toprule
$y$ & $K^+$ & $\Delta K^+$ & $K^-$ & $\Delta K^-$ & $K^{*0}$ & $\Delta K^{*0}$ & $\bar{K}^{*0}$ & $\Delta \bar{K}^{*0}$ \\
\midrule
\endfirsthead

\multicolumn{9}{c}{\tablename\ \thetable\ -- continued from previous page} \\
\toprule
$y$ & $K^+$ & $\Delta K^+$ & $K^-$ & $\Delta K^-$ & $K^{*0}$ & $\Delta K^{*0}$ & $\bar{K}^{*0}$ & $\Delta \bar{K}^{*0}$ \\
\midrule
\endhead

\midrule
\multicolumn{9}{r}{continued on next page} \\
\endfoot

\bottomrule
\endlastfoot

-3.9500 & 8.0000e-05 & 4.0000e-05 & 2.0000e-05 & 2.0000e-05 & 0.0000e+00 & 0.0000e+00 & 0.0000e+00 & 0.0000e+00 \\
-3.8500 & 1.0000e-04 & 4.4721e-05 & 0.0000e+00 & 0.0000e+00 & 0.0000e+00 & 0.0000e+00 & 0.0000e+00 & 0.0000e+00 \\
-3.7500 & 2.6000e-04 & 7.2111e-05 & 1.0000e-04 & 4.4721e-05 & 0.0000e+00 & 0.0000e+00 & 0.0000e+00 & 0.0000e+00 \\
-3.6500 & 1.0000e-03 & 1.4142e-04 & 2.0000e-04 & 6.3246e-05 & 0.0000e+00 & 0.0000e+00 & 0.0000e+00 & 0.0000e+00 \\
-3.5500 & 2.2000e-03 & 2.0976e-04 & 4.8000e-04 & 9.7980e-05 & 0.0000e+00 & 0.0000e+00 & 0.0000e+00 & 0.0000e+00 \\
-3.4500 & 4.8400e-03 & 3.1113e-04 & 8.4000e-04 & 1.2961e-04 & 4.0000e-05 & 2.8284e-05 & 0.0000e+00 & 0.0000e+00 \\
-3.3500 & 1.0960e-02 & 4.6819e-04 & 2.1000e-03 & 2.0494e-04 & 1.4000e-04 & 5.2915e-05 & 2.0000e-05 & 2.0000e-05 \\
-3.2500 & 2.1540e-02 & 6.5635e-04 & 4.7800e-03 & 3.0919e-04 & 3.2000e-04 & 8.0000e-05 & 1.0000e-04 & 4.4721e-05 \\
-3.1500 & 4.1940e-02 & 9.1586e-04 & 9.0200e-03 & 4.2474e-04 & 1.1200e-03 & 1.4967e-04 & 2.2000e-04 & 6.6332e-05 \\
-3.0500 & 7.1120e-02 & 1.1926e-03 & 1.4740e-02 & 5.4295e-04 & 2.9000e-03 & 2.4083e-04 & 6.0000e-04 & 1.0954e-04 \\
-2.9500 & 1.1808e-01 & 1.5367e-03 & 2.6860e-02 & 7.3294e-04 & 5.8000e-03 & 3.4059e-04 & 1.4600e-03 & 1.7088e-04 \\
-2.8500 & 1.7206e-01 & 1.8550e-03 & 4.0800e-02 & 9.0333e-04 & 1.2800e-02 & 5.0596e-04 & 2.1800e-03 & 2.0881e-04 \\
-2.7500 & 2.5822e-01 & 2.2725e-03 & 6.4960e-02 & 1.1398e-03 & 2.5680e-02 & 7.1666e-04 & 4.5400e-03 & 3.0133e-04 \\
-2.6500 & 3.5386e-01 & 2.6603e-03 & 9.5020e-02 & 1.3785e-03 & 4.7900e-02 & 9.7877e-04 & 8.7400e-03 & 4.1809e-04 \\
-2.5500 & 4.6660e-01 & 3.0548e-03 & 1.3328e-01 & 1.6327e-03 & 7.6980e-02 & 1.2408e-03 & 1.4960e-02 & 5.4699e-04 \\
-2.4500 & 6.0570e-01 & 3.4805e-03 & 1.8282e-01 & 1.9122e-03 & 1.1352e-01 & 1.5068e-03 & 2.2960e-02 & 6.7764e-04 \\
-2.3500 & 7.6114e-01 & 3.9016e-03 & 2.4138e-01 & 2.1972e-03 & 1.6526e-01 & 1.8180e-03 & 3.4520e-02 & 8.3090e-04 \\
-2.2500 & 9.2870e-01 & 4.3098e-03 & 3.2258e-01 & 2.5400e-03 & 2.1364e-01 & 2.0671e-03 & 4.7700e-02 & 9.7673e-04 \\
-2.1500 & 1.1016e+00 & 4.6938e-03 & 4.1532e-01 & 2.8821e-03 & 2.7456e-01 & 2.3433e-03 & 6.8920e-02 & 1.1741e-03 \\
-2.0500 & 1.2731e+00 & 5.0460e-03 & 5.1408e-01 & 3.2065e-03 & 3.2780e-01 & 2.5605e-03 & 9.0540e-02 & 1.3457e-03 \\
-1.9500 & 1.4390e+00 & 5.3647e-03 & 6.3260e-01 & 3.5570e-03 & 3.7946e-01 & 2.7549e-03 & 1.1642e-01 & 1.5259e-03 \\
-1.8500 & 1.6075e+00 & 5.6700e-03 & 7.5466e-01 & 3.8850e-03 & 4.3682e-01 & 2.9557e-03 & 1.4566e-01 & 1.7068e-03 \\
-1.7500 & 1.7653e+00 & 5.9419e-03 & 8.9154e-01 & 4.2227e-03 & 5.0074e-01 & 3.1646e-03 & 1.7732e-01 & 1.8832e-03 \\
-1.6500 & 1.9254e+00 & 6.2055e-03 & 1.0219e+00 & 4.5209e-03 & 5.3864e-01 & 3.2822e-03 & 2.1334e-01 & 2.0656e-03 \\
-1.5500 & 2.0686e+00 & 6.4320e-03 & 1.1717e+00 & 4.8409e-03 & 5.9414e-01 & 3.4471e-03 & 2.4472e-01 & 2.2123e-03 \\
-1.4500 & 2.2086e+00 & 6.6461e-03 & 1.3165e+00 & 5.1312e-03 & 6.3608e-01 & 3.5667e-03 & 2.8386e-01 & 2.3827e-03 \\
-1.3500 & 2.3510e+00 & 6.8571e-03 & 1.4565e+00 & 5.3972e-03 & 6.7790e-01 & 3.6821e-03 & 3.2528e-01 & 2.5506e-03 \\
-1.2500 & 2.4805e+00 & 7.0435e-03 & 1.6029e+00 & 5.6620e-03 & 7.1952e-01 & 3.7935e-03 & 3.6256e-01 & 2.6928e-03 \\
-1.1500 & 2.5931e+00 & 7.2015e-03 & 1.7472e+00 & 5.9113e-03 & 7.5520e-01 & 3.8864e-03 & 3.9606e-01 & 2.8145e-03 \\
-1.0500 & 2.6986e+00 & 7.3466e-03 & 1.8691e+00 & 6.1140e-03 & 7.9424e-01 & 3.9856e-03 & 4.3626e-01 & 2.9538e-03 \\
-0.9500 & 2.8039e+00 & 7.4885e-03 & 1.9766e+00 & 6.2874e-03 & 8.2684e-01 & 4.0665e-03 & 4.6592e-01 & 3.0526e-03 \\
-0.8500 & 2.8991e+00 & 7.6146e-03 & 2.0852e+00 & 6.4578e-03 & 8.5452e-01 & 4.1341e-03 & 4.9604e-01 & 3.1497e-03 \\
-0.7500 & 2.9816e+00 & 7.7222e-03 & 2.1869e+00 & 6.6135e-03 & 8.8330e-01 & 4.2031e-03 & 5.2420e-01 & 3.2379e-03 \\
-0.6500 & 3.0569e+00 & 7.8190e-03 & 2.2681e+00 & 6.7352e-03 & 9.0870e-01 & 4.2631e-03 & 5.4696e-01 & 3.3074e-03 \\
-0.5500 & 3.1066e+00 & 7.8824e-03 & 2.3407e+00 & 6.8421e-03 & 9.3002e-01 & 4.3128e-03 & 5.7524e-01 & 3.3919e-03 \\
-0.4500 & 3.1724e+00 & 7.9655e-03 & 2.4167e+00 & 6.9522e-03 & 9.3638e-01 & 4.3275e-03 & 5.9840e-01 & 3.4595e-03 \\
-0.3500 & 3.2100e+00 & 8.0124e-03 & 2.4468e+00 & 6.9955e-03 & 9.5694e-01 & 4.3748e-03 & 6.1132e-01 & 3.4966e-03 \\
-0.2500 & 3.2153e+00 & 8.0191e-03 & 2.4755e+00 & 7.0363e-03 & 9.6548e-01 & 4.3943e-03 & 6.2696e-01 & 3.5411e-03 \\
-0.1500 & 3.2656e+00 & 8.0816e-03 & 2.4954e+00 & 7.0646e-03 & 9.7796e-01 & 4.4226e-03 & 6.3110e-01 & 3.5527e-03 \\
-0.0500 & 3.2679e+00 & 8.0845e-03 & 2.5084e+00 & 7.0830e-03 & 9.7448e-01 & 4.4147e-03 & 6.3652e-01 & 3.5680e-03 \\
0.0500 & 3.2553e+00 & 8.0689e-03 & 2.5141e+00 & 7.0910e-03 & 9.7502e-01 & 4.4159e-03 & 6.3292e-01 & 3.5579e-03 \\
0.1500 & 3.2322e+00 & 8.0401e-03 & 2.4882e+00 & 7.0543e-03 & 9.7706e-01 & 4.4205e-03 & 6.3106e-01 & 3.5526e-03 \\
0.2500 & 3.2076e+00 & 8.0095e-03 & 2.4695e+00 & 7.0279e-03 & 9.5868e-01 & 4.3788e-03 & 6.1874e-01 & 3.5178e-03 \\
0.3500 & 3.1608e+00 & 7.9508e-03 & 2.4291e+00 & 6.9701e-03 & 9.5246e-01 & 4.3645e-03 & 6.0632e-01 & 3.4823e-03 \\
0.4500 & 3.1195e+00 & 7.8988e-03 & 2.3778e+00 & 6.8960e-03 & 9.3406e-01 & 4.3222e-03 & 5.9600e-01 & 3.4525e-03 \\
0.5500 & 3.0639e+00 & 7.8281e-03 & 2.3017e+00 & 6.7849e-03 & 9.1300e-01 & 4.2732e-03 & 5.6896e-01 & 3.3733e-03 \\
0.6500 & 2.9874e+00 & 7.7297e-03 & 2.2469e+00 & 6.7035e-03 & 8.8928e-01 & 4.2173e-03 & 5.4878e-01 & 3.3129e-03 \\
0.7500 & 2.8949e+00 & 7.6091e-03 & 2.1461e+00 & 6.5514e-03 & 8.7062e-01 & 4.1728e-03 & 5.1792e-01 & 3.2184e-03 \\
0.8500 & 2.8081e+00 & 7.4941e-03 & 2.0448e+00 & 6.3950e-03 & 8.3910e-01 & 4.0966e-03 & 4.9344e-01 & 3.1415e-03 \\
0.9500 & 2.7238e+00 & 7.3808e-03 & 1.9389e+00 & 6.2272e-03 & 8.0156e-01 & 4.0039e-03 & 4.6096e-01 & 3.0363e-03 \\
1.0500 & 2.6201e+00 & 7.2390e-03 & 1.8199e+00 & 6.0330e-03 & 7.6688e-01 & 3.9163e-03 & 4.2574e-01 & 2.9180e-03 \\
1.1500 & 2.4870e+00 & 7.0527e-03 & 1.6911e+00 & 5.8157e-03 & 7.2790e-01 & 3.8155e-03 & 3.9134e-01 & 2.7976e-03 \\
1.2500 & 2.3659e+00 & 6.8788e-03 & 1.5494e+00 & 5.5667e-03 & 6.9734e-01 & 3.7345e-03 & 3.5244e-01 & 2.6550e-03 \\
1.3500 & 2.2478e+00 & 6.7049e-03 & 1.4055e+00 & 5.3020e-03 & 6.5412e-01 & 3.6170e-03 & 3.1322e-01 & 2.5029e-03 \\
1.4500 & 2.1071e+00 & 6.4917e-03 & 1.2616e+00 & 5.0231e-03 & 6.0920e-01 & 3.4906e-03 & 2.7838e-01 & 2.3596e-03 \\
1.5500 & 1.9739e+00 & 6.2832e-03 & 1.1253e+00 & 4.7440e-03 & 5.6316e-01 & 3.3561e-03 & 2.4058e-01 & 2.1935e-03 \\
1.6500 & 1.8173e+00 & 6.0288e-03 & 9.8314e-01 & 4.4343e-03 & 5.1332e-01 & 3.2041e-03 & 2.0470e-01 & 2.0234e-03 \\
1.7500 & 1.6789e+00 & 5.7946e-03 & 8.5424e-01 & 4.1334e-03 & 4.6586e-01 & 3.0524e-03 & 1.7180e-01 & 1.8536e-03 \\
1.8500 & 1.5162e+00 & 5.5068e-03 & 7.1586e-01 & 3.7838e-03 & 4.0986e-01 & 2.8631e-03 & 1.4310e-01 & 1.6917e-03 \\
1.9500 & 1.3591e+00 & 5.2136e-03 & 5.9746e-01 & 3.4568e-03 & 3.7122e-01 & 2.7248e-03 & 1.0972e-01 & 1.4814e-03 \\
2.0500 & 1.1849e+00 & 4.8681e-03 & 4.8108e-01 & 3.1019e-03 & 3.0808e-01 & 2.4823e-03 & 8.7360e-02 & 1.3218e-03 \\
2.1500 & 1.0261e+00 & 4.5302e-03 & 3.8242e-01 & 2.7656e-03 & 2.5568e-01 & 2.2613e-03 & 6.5320e-02 & 1.1430e-03 \\
2.2500 & 8.5416e-01 & 4.1332e-03 & 2.9680e-01 & 2.4364e-03 & 1.9824e-01 & 1.9912e-03 & 4.4720e-02 & 9.4573e-04 \\
2.3500 & 6.9870e-01 & 3.7382e-03 & 2.2190e-01 & 2.1067e-03 & 1.4860e-01 & 1.7239e-03 & 3.1180e-02 & 7.8968e-04 \\
2.4500 & 5.6056e-01 & 3.3483e-03 & 1.6538e-01 & 1.8187e-03 & 1.0592e-01 & 1.4555e-03 & 2.0660e-02 & 6.4281e-04 \\
2.5500 & 4.2744e-01 & 2.9238e-03 & 1.1730e-01 & 1.5317e-03 & 7.0920e-02 & 1.1910e-03 & 1.1880e-02 & 4.8744e-04 \\
2.6500 & 3.1644e-01 & 2.5157e-03 & 8.0460e-02 & 1.2685e-03 & 3.9960e-02 & 8.9398e-04 & 8.1800e-03 & 4.0447e-04 \\
2.7500 & 2.2808e-01 & 2.1358e-03 & 5.4260e-02 & 1.0417e-03 & 2.3400e-02 & 6.8411e-04 & 4.2400e-03 & 2.9120e-04 \\
2.8500 & 1.5598e-01 & 1.7662e-03 & 3.5100e-02 & 8.3785e-04 & 1.1060e-02 & 4.7032e-04 & 1.9000e-03 & 1.9494e-04 \\
2.9500 & 9.9900e-02 & 1.4135e-03 & 2.2920e-02 & 6.7705e-04 & 5.0600e-03 & 3.1812e-04 & 1.0600e-03 & 1.4560e-04 \\
3.0500 & 6.1760e-02 & 1.1114e-03 & 1.2600e-02 & 5.0200e-04 & 2.2800e-03 & 2.1354e-04 & 4.8000e-04 & 9.7980e-05 \\
3.1500 & 3.5360e-02 & 8.4095e-04 & 7.6600e-03 & 3.9141e-04 & 6.6000e-04 & 1.1489e-04 & 1.0000e-04 & 4.4721e-05 \\
3.2500 & 1.8620e-02 & 6.1025e-04 & 3.7400e-03 & 2.7350e-04 & 2.4000e-04 & 6.9282e-05 & 2.0000e-05 & 2.0000e-05 \\
3.3500 & 8.0200e-03 & 4.0050e-04 & 2.0800e-03 & 2.0396e-04 & 6.0000e-05 & 3.4641e-05 & 0.0000e+00 & 0.0000e+00 \\
3.4500 & 3.6200e-03 & 2.6907e-04 & 1.1400e-03 & 1.5100e-04 & 2.0000e-05 & 2.0000e-05 & 0.0000e+00 & 0.0000e+00 \\
3.5500 & 1.1600e-03 & 1.5232e-04 & 3.4000e-04 & 8.2462e-05 & 2.0000e-05 & 2.0000e-05 & 0.0000e+00 & 0.0000e+00 \\
3.6500 & 7.2000e-04 & 1.2000e-04 & 1.8000e-04 & 6.0000e-05 & 0.0000e+00 & 0.0000e+00 & 0.0000e+00 & 0.0000e+00 \\
3.7500 & 2.4000e-04 & 6.9282e-05 & 6.0000e-05 & 3.4641e-05 & 0.0000e+00 & 0.0000e+00 & 0.0000e+00 & 0.0000e+00 \\
3.8500 & 6.0000e-05 & 3.4641e-05 & 2.0000e-05 & 2.0000e-05 & 0.0000e+00 & 0.0000e+00 & 0.0000e+00 & 0.0000e+00 \\
3.9500 & 2.0000e-05 & 2.0000e-05 & 2.0000e-05 & 2.0000e-05 & 0.0000e+00 & 0.0000e+00 & 0.0000e+00 & 0.0000e+00 \\
\end{longtable}

\subsection{Numerical data used in Fig.~\ref{fig:plot_pT_dNdpT} for $\sqrt{s_{NN}}=8.8, 11.9, 16.8$~GeV}

\scriptsize
\setlength{\LTleft}{0pt}
\setlength{\LTright}{0pt}

\begin{longtable}{c c c c c c c c c}
\caption{Numerical values used for the UrQMD transverse momentum distributions in Ar+Sc collisions at $\sqrt{s_{NN}}=8.8$~GeV. The table lists the transverse momentum bin center $p_T$, the yields, and the corresponding statistical uncertainties for $K^+$, $K^-$, $K^{*0}$, and $\bar{K}^{*0}$.}
\label{tab:dndpt_ecm8p8}\\
\toprule
$p_T$ & $K^+$ & $\Delta K^+$ & $K^-$ & $\Delta K^-$ & $K^{*0}$ & $\Delta K^{*0}$ & $\bar{K}^{*0}$ & $\Delta \bar{K}^{*0}$ \\
\midrule
\endfirsthead

\multicolumn{9}{c}{\tablename\ \thetable\ -- continued from previous page} \\
\toprule
$p_T$ & $K^+$ & $\Delta K^+$ & $K^-$ & $\Delta K^-$ & $K^{*0}$ & $\Delta K^{*0}$ & $\bar{K}^{*0}$ & $\Delta \bar{K}^{*0}$ \\
\midrule
\endhead

\midrule
\multicolumn{9}{r}{continued on next page} \\
\endfoot

\bottomrule
\endlastfoot

0.0500 & 7.6933e-01 & 3.2028e-03 & 4.7305e-01 & 2.5114e-03 & 1.0883e-01 & 1.2046e-03 & 3.8893e-02 & 7.2012e-04 \\
0.1500 & 2.0715e+00 & 5.2554e-03 & 1.2710e+00 & 4.1167e-03 & 3.0639e-01 & 2.0212e-03 & 1.0987e-01 & 1.2103e-03 \\
0.2500 & 2.8151e+00 & 6.1266e-03 & 1.6877e+00 & 4.7437e-03 & 4.7588e-01 & 2.5189e-03 & 1.6877e-01 & 1.5001e-03 \\
0.3500 & 2.9335e+00 & 6.2541e-03 & 1.7130e+00 & 4.7791e-03 & 5.8289e-01 & 2.7878e-03 & 2.0948e-01 & 1.6712e-03 \\
0.4500 & 2.6091e+00 & 5.8981e-03 & 1.4599e+00 & 4.4120e-03 & 6.2839e-01 & 2.8946e-03 & 2.3408e-01 & 1.7667e-03 \\
0.5500 & 2.1075e+00 & 5.3009e-03 & 1.1116e+00 & 3.8498e-03 & 6.1559e-01 & 2.8649e-03 & 2.2905e-01 & 1.7476e-03 \\
0.6500 & 1.6080e+00 & 4.6304e-03 & 8.0172e-01 & 3.2695e-03 & 5.5529e-01 & 2.7210e-03 & 2.1280e-01 & 1.6844e-03 \\
0.7500 & 1.1682e+00 & 3.9466e-03 & 5.6141e-01 & 2.7360e-03 & 4.7277e-01 & 2.5107e-03 & 1.8359e-01 & 1.5646e-03 \\
0.8500 & 8.3075e-01 & 3.3282e-03 & 3.8172e-01 & 2.2560e-03 & 3.7560e-01 & 2.2379e-03 & 1.5289e-01 & 1.4278e-03 \\
0.9500 & 5.7901e-01 & 2.7785e-03 & 2.5427e-01 & 1.8413e-03 & 2.9669e-01 & 1.9889e-03 & 1.1709e-01 & 1.2495e-03 \\
1.0500 & 3.9823e-01 & 2.3043e-03 & 1.6765e-01 & 1.4951e-03 & 2.2319e-01 & 1.7251e-03 & 8.8213e-02 & 1.0845e-03 \\
1.1500 & 2.7436e-01 & 1.9126e-03 & 1.1311e-01 & 1.2280e-03 & 1.6685e-01 & 1.4915e-03 & 6.4640e-02 & 9.2837e-04 \\
1.2500 & 1.8881e-01 & 1.5867e-03 & 7.4507e-02 & 9.9671e-04 & 1.1988e-01 & 1.2643e-03 & 4.6453e-02 & 7.8701e-04 \\
1.3500 & 1.2957e-01 & 1.3144e-03 & 4.9080e-02 & 8.0895e-04 & 8.3600e-02 & 1.0558e-03 & 3.2720e-02 & 6.6050e-04 \\
1.4500 & 9.0360e-02 & 1.0976e-03 & 3.1573e-02 & 6.4883e-04 & 6.0920e-02 & 9.0126e-04 & 2.2853e-02 & 5.5201e-04 \\
1.5500 & 6.0853e-02 & 9.0077e-04 & 2.1667e-02 & 5.3748e-04 & 4.2467e-02 & 7.5248e-04 & 1.5787e-02 & 4.5879e-04 \\
1.6500 & 4.1987e-02 & 7.4821e-04 & 1.4080e-02 & 4.3328e-04 & 2.9240e-02 & 6.2439e-04 & 1.0560e-02 & 3.7523e-04 \\
1.7500 & 2.9147e-02 & 6.2340e-04 & 8.5733e-03 & 3.3810e-04 & 2.1253e-02 & 5.3233e-04 & 6.9333e-03 & 3.0405e-04 \\
1.8500 & 2.0000e-02 & 5.1640e-04 & 5.8800e-03 & 2.8000e-04 & 1.4600e-02 & 4.4121e-04 & 4.5733e-03 & 2.4694e-04 \\
1.9500 & 1.3107e-02 & 4.1804e-04 & 4.1467e-03 & 2.3514e-04 & 1.0013e-02 & 3.6539e-04 & 3.3067e-03 & 2.0997e-04 \\
2.0500 & 9.8800e-03 & 3.6295e-04 & 2.5333e-03 & 1.8379e-04 & 7.4667e-03 & 3.1552e-04 & 2.2133e-03 & 1.7179e-04 \\
2.1500 & 6.2800e-03 & 2.8937e-04 & 1.8267e-03 & 1.5606e-04 & 5.3200e-03 & 2.6633e-04 & 1.4533e-03 & 1.3920e-04 \\
2.2500 & 5.0000e-03 & 2.5820e-04 & 1.0133e-03 & 1.1624e-04 & 3.7733e-03 & 2.2430e-04 & 9.0667e-04 & 1.0995e-04 \\
2.3500 & 3.2800e-03 & 2.0913e-04 & 7.4667e-04 & 9.9778e-05 & 2.3467e-03 & 1.7689e-04 & 7.0667e-04 & 9.7068e-05 \\
2.4500 & 2.6400e-03 & 1.8762e-04 & 5.6000e-04 & 8.6410e-05 & 1.5467e-03 & 1.4360e-04 & 4.2667e-04 & 7.5425e-05 \\
2.5500 & 1.6533e-03 & 1.4847e-04 & 4.9333e-04 & 8.1104e-05 & 1.2533e-03 & 1.2927e-04 & 2.8000e-04 & 6.1101e-05 \\
2.6500 & 1.2267e-03 & 1.2789e-04 & 1.8667e-04 & 4.9889e-05 & 9.4667e-04 & 1.1235e-04 & 1.7333e-04 & 4.8074e-05 \\
2.7500 & 7.7333e-04 & 1.0154e-04 & 1.3333e-04 & 4.2164e-05 & 5.3333e-04 & 8.4327e-05 & 8.0000e-05 & 3.2660e-05 \\
2.8500 & 4.2667e-04 & 7.5425e-05 & 1.0667e-04 & 3.7712e-05 & 3.2000e-04 & 6.5320e-05 & 1.4667e-04 & 4.4222e-05 \\
2.9500 & 2.9333e-04 & 6.2539e-05 & 8.0000e-05 & 3.2660e-05 & 2.8000e-04 & 6.1101e-05 & 0.0000e+00 & 0.0000e+00 \\
\end{longtable}

\scriptsize
\setlength{\LTleft}{0pt}
\setlength{\LTright}{0pt}

\begin{longtable}{c c c c c c c c c}
\caption{Numerical values used for the UrQMD transverse momentum distributions in Ar+Sc collisions at $\sqrt{s_{NN}}=11.9$~GeV. The table lists the transverse momentum bin center $p_T$, the yields, and the corresponding statistical uncertainties for $K^+$, $K^-$, $K^{*0}$, and $\bar{K}^{*0}$.}
\label{tab:dndpt_ecm8p8}\\
\toprule
$p_T$ & $K^+$ & $\Delta K^+$ & $K^-$ & $\Delta K^-$ & $K^{*0}$ & $\Delta K^{*0}$ & $\bar{K}^{*0}$ & $\Delta \bar{K}^{*0}$ \\
\midrule
\endfirsthead

\multicolumn{9}{c}{\tablename\ \thetable\ -- continued from previous page} \\
\toprule
$p_T$ & $K^+$ & $\Delta K^+$ & $K^-$ & $\Delta K^-$ & $K^{*0}$ & $\Delta K^{*0}$ & $\bar{K}^{*0}$ & $\Delta \bar{K}^{*0}$ \\
\midrule
\endhead

\midrule
\multicolumn{9}{r}{continued on next page} \\
\endfoot

\bottomrule
\endlastfoot

0.0500 & 9.4903e-01 & 4.0644e-03 & 6.5706e-01 & 3.3819e-03 & 1.4132e-01 & 1.5684e-03 & 6.3011e-02 & 1.0473e-03 \\
0.1500 & 2.5708e+00 & 6.6894e-03 & 1.7817e+00 & 5.5689e-03 & 4.0028e-01 & 2.6396e-03 & 1.8416e-01 & 1.7904e-03 \\
0.2500 & 3.4580e+00 & 7.7584e-03 & 2.4013e+00 & 6.4651e-03 & 6.0874e-01 & 3.2551e-03 & 2.8125e-01 & 2.2126e-03 \\
0.3500 & 3.6370e+00 & 7.9566e-03 & 2.4713e+00 & 6.5587e-03 & 7.5911e-01 & 3.6350e-03 & 3.5469e-01 & 2.4847e-03 \\
0.4500 & 3.2608e+00 & 7.5338e-03 & 2.1446e+00 & 6.1098e-03 & 8.1636e-01 & 3.7696e-03 & 3.9248e-01 & 2.6137e-03 \\
0.5500 & 2.6375e+00 & 6.7757e-03 & 1.6519e+00 & 5.3623e-03 & 7.9645e-01 & 3.7234e-03 & 3.8754e-01 & 2.5972e-03 \\
0.6500 & 1.9953e+00 & 5.8933e-03 & 1.2092e+00 & 4.5878e-03 & 7.1605e-01 & 3.5304e-03 & 3.5598e-01 & 2.4892e-03 \\
0.7500 & 1.4663e+00 & 5.0520e-03 & 8.5652e-01 & 3.8612e-03 & 6.1105e-01 & 3.2613e-03 & 3.0778e-01 & 2.3146e-03 \\
0.8500 & 1.0330e+00 & 4.2404e-03 & 5.8265e-01 & 3.1846e-03 & 4.9788e-01 & 2.9439e-03 & 2.5144e-01 & 2.0920e-03 \\
0.9500 & 7.1943e-01 & 3.5387e-03 & 3.9814e-01 & 2.6325e-03 & 3.8498e-01 & 2.5886e-03 & 1.9586e-01 & 1.8464e-03 \\
1.0500 & 5.0326e-01 & 2.9597e-03 & 2.6693e-01 & 2.1555e-03 & 2.9163e-01 & 2.2530e-03 & 1.4597e-01 & 1.5940e-03 \\
1.1500 & 3.4616e-01 & 2.4547e-03 & 1.7568e-01 & 1.7487e-03 & 2.1807e-01 & 1.9483e-03 & 1.1225e-01 & 1.3978e-03 \\
1.2500 & 2.3283e-01 & 2.0131e-03 & 1.1647e-01 & 1.4238e-03 & 1.5836e-01 & 1.6603e-03 & 8.0244e-02 & 1.1818e-03 \\
1.3500 & 1.6003e-01 & 1.6690e-03 & 7.8451e-02 & 1.1686e-03 & 1.1288e-01 & 1.4017e-03 & 5.5544e-02 & 9.8327e-04 \\
1.4500 & 1.0832e-01 & 1.3731e-03 & 5.1436e-02 & 9.4621e-04 & 7.9095e-02 & 1.1734e-03 & 3.8625e-02 & 8.1995e-04 \\
1.5500 & 7.4395e-02 & 1.1380e-03 & 3.4064e-02 & 7.7003e-04 & 5.5074e-02 & 9.7910e-04 & 2.6040e-02 & 6.7325e-04 \\
1.6500 & 5.0688e-02 & 9.3930e-04 & 2.2785e-02 & 6.2977e-04 & 3.8242e-02 & 8.1588e-04 & 1.8695e-02 & 5.7044e-04 \\
1.7500 & 3.4378e-02 & 7.7356e-04 & 1.5579e-02 & 5.2074e-04 & 2.6545e-02 & 6.7974e-04 & 1.1715e-02 & 4.5156e-04 \\
1.8500 & 2.5466e-02 & 6.6578e-04 & 9.1558e-03 & 3.9921e-04 & 1.8416e-02 & 5.6618e-04 & 8.4073e-03 & 3.8255e-04 \\
1.9500 & 1.7198e-02 & 5.4713e-04 & 6.3534e-03 & 3.3255e-04 & 1.3734e-02 & 4.8893e-04 & 5.2567e-03 & 3.0249e-04 \\
2.0500 & 1.1941e-02 & 4.5590e-04 & 4.1253e-03 & 2.6797e-04 & 9.5213e-03 & 4.0710e-04 & 3.8816e-03 & 2.5993e-04 \\
2.1500 & 7.9025e-03 & 3.7088e-04 & 2.9417e-03 & 2.2628e-04 & 6.4752e-03 & 3.3572e-04 & 2.5065e-03 & 2.0888e-04 \\
2.2500 & 5.6571e-03 & 3.1380e-04 & 2.2802e-03 & 1.9923e-04 & 4.3168e-03 & 2.7412e-04 & 1.6188e-03 & 1.6786e-04 \\
2.3500 & 3.9339e-03 & 2.6168e-04 & 1.4273e-03 & 1.5762e-04 & 3.3246e-03 & 2.4056e-04 & 1.2010e-03 & 1.4459e-04 \\
2.4500 & 2.9417e-03 & 2.2628e-04 & 8.7032e-04 & 1.2308e-04 & 2.1410e-03 & 1.9305e-04 & 1.0270e-03 & 1.3370e-04 \\
2.5500 & 2.1410e-03 & 1.9305e-04 & 5.9182e-04 & 1.0150e-04 & 1.5492e-03 & 1.6421e-04 & 4.5257e-04 & 8.8756e-05 \\
2.6500 & 1.4273e-03 & 1.5762e-04 & 3.4813e-04 & 7.7844e-05 & 1.1662e-03 & 1.4248e-04 & 4.3516e-04 & 8.7032e-05 \\
2.7500 & 7.6588e-04 & 1.1546e-04 & 2.0888e-04 & 6.0298e-05 & 7.3107e-04 & 1.1281e-04 & 1.3925e-04 & 4.9233e-05 \\
2.8500 & 7.4848e-04 & 1.1414e-04 & 1.2185e-04 & 4.6053e-05 & 5.5701e-04 & 9.8466e-05 & 2.4369e-04 & 6.5129e-05 \\
2.9500 & 5.5701e-04 & 9.8466e-05 & 1.3925e-04 & 4.9233e-05 & 4.1775e-04 & 8.5274e-05 & 6.9626e-05 & 3.4813e-05 \\
\end{longtable}

\scriptsize
\setlength{\LTleft}{0pt}
\setlength{\LTright}{0pt}

\begin{longtable}{c c c c c c c c c}
\caption{Numerical values used for the UrQMD transverse momentum distributions in Ar+Sc collisions at $\sqrt{s_{NN}}=16.8$~GeV. The table lists the transverse momentum bin center $p_T$, the yields, and the corresponding statistical uncertainties for $K^+$, $K^-$, $K^{*0}$, and $\bar{K}^{*0}$.}
\label{tab:dndpt_ecm8p8}\\
\toprule
$p_T$ & $K^+$ & $\Delta K^+$ & $K^-$ & $\Delta K^-$ & $K^{*0}$ & $\Delta K^{*0}$ & $\bar{K}^{*0}$ & $\Delta \bar{K}^{*0}$ \\
\midrule
\endfirsthead

\multicolumn{9}{c}{\tablename\ \thetable\ -- continued from previous page} \\
\toprule
$p_T$ & $K^+$ & $\Delta K^+$ & $K^-$ & $\Delta K^-$ & $K^{*0}$ & $\Delta K^{*0}$ & $\bar{K}^{*0}$ & $\Delta \bar{K}^{*0}$ \\
\midrule
\endhead

\midrule
\multicolumn{9}{r}{continued on next page} \\
\endfoot

\bottomrule
\endlastfoot

0.0500 & 1.1619e+00 & 4.6193e-03 & 8.8481e-01 & 4.0311e-03 & 1.7719e-01 & 1.8039e-03 & 9.9853e-02 & 1.3542e-03 \\
0.1500 & 3.1281e+00 & 7.5795e-03 & 2.3898e+00 & 6.6250e-03 & 5.1106e-01 & 3.0636e-03 & 2.8118e-01 & 2.2724e-03 \\
0.2500 & 4.2470e+00 & 8.8317e-03 & 3.2391e+00 & 7.7128e-03 & 7.8621e-01 & 3.7999e-03 & 4.4017e-01 & 2.8432e-03 \\
0.3500 & 4.4448e+00 & 9.0350e-03 & 3.3518e+00 & 7.8459e-03 & 9.6048e-01 & 4.2000e-03 & 5.5012e-01 & 3.1786e-03 \\
0.4500 & 3.9919e+00 & 8.5623e-03 & 2.9250e+00 & 7.3293e-03 & 1.0437e+00 & 4.3780e-03 & 6.0312e-01 & 3.3282e-03 \\
0.5500 & 3.2424e+00 & 7.7168e-03 & 2.3017e+00 & 6.5016e-03 & 1.0207e+00 & 4.3296e-03 & 6.0184e-01 & 3.3246e-03 \\
0.6500 & 2.4548e+00 & 6.7145e-03 & 1.7098e+00 & 5.6036e-03 & 9.1706e-01 & 4.1039e-03 & 5.4896e-01 & 3.1752e-03 \\
0.7500 & 1.7769e+00 & 5.7126e-03 & 1.2186e+00 & 4.7308e-03 & 7.6955e-01 & 3.7594e-03 & 4.7517e-01 & 2.9541e-03 \\
0.8500 & 1.2582e+00 & 4.8070e-03 & 8.4762e-01 & 3.9455e-03 & 6.3069e-01 & 3.4034e-03 & 3.8896e-01 & 2.6727e-03 \\
0.9500 & 8.7298e-01 & 4.0041e-03 & 5.6966e-01 & 3.2345e-03 & 4.8538e-01 & 2.9857e-03 & 3.0823e-01 & 2.3792e-03 \\
1.0500 & 6.0185e-01 & 3.3247e-03 & 3.8747e-01 & 2.6676e-03 & 3.7212e-01 & 2.6142e-03 & 2.2797e-01 & 2.0462e-03 \\
1.1500 & 4.1572e-01 & 2.7631e-03 & 2.5809e-01 & 2.1771e-03 & 2.7223e-01 & 2.2360e-03 & 1.6940e-01 & 1.7639e-03 \\
1.2500 & 2.7770e-01 & 2.2584e-03 & 1.7230e-01 & 1.7789e-03 & 1.9596e-01 & 1.8971e-03 & 1.2015e-01 & 1.4854e-03 \\
1.3500 & 1.8795e-01 & 1.8579e-03 & 1.1499e-01 & 1.4532e-03 & 1.3846e-01 & 1.5946e-03 & 8.5565e-02 & 1.2536e-03 \\
1.4500 & 1.2865e-01 & 1.5371e-03 & 7.6657e-02 & 1.1865e-03 & 9.5152e-02 & 1.3219e-03 & 5.8953e-02 & 1.0405e-03 \\
1.5500 & 8.9587e-02 & 1.2827e-03 & 5.0634e-02 & 9.6432e-04 & 6.8118e-02 & 1.1185e-03 & 4.0367e-02 & 8.6103e-04 \\
1.6500 & 6.1983e-02 & 1.0669e-03 & 3.3480e-02 & 7.8414e-04 & 4.8430e-02 & 9.4310e-04 & 2.7420e-02 & 7.0963e-04 \\
1.7500 & 4.2112e-02 & 8.7944e-04 & 2.2810e-02 & 6.4724e-04 & 3.3903e-02 & 7.8907e-04 & 1.8567e-02 & 5.8395e-04 \\
1.8500 & 2.7420e-02 & 7.0963e-04 & 1.5262e-02 & 5.2942e-04 & 2.3857e-02 & 6.6192e-04 & 1.3205e-02 & 4.9246e-04 \\
1.9500 & 2.1524e-02 & 6.2873e-04 & 1.0762e-02 & 4.4458e-04 & 1.5776e-02 & 5.3827e-04 & 8.9807e-03 & 4.0612e-04 \\
2.0500 & 1.4362e-02 & 5.1358e-04 & 7.2544e-03 & 3.6501e-04 & 1.1644e-02 & 4.6243e-04 & 6.0055e-03 & 3.3211e-04 \\
2.1500 & 1.0101e-02 & 4.3071e-04 & 4.1873e-03 & 2.7731e-04 & 7.9155e-03 & 3.8128e-04 & 4.3710e-03 & 2.8333e-04 \\
2.2500 & 7.4013e-03 & 3.6868e-04 & 3.0119e-03 & 2.3519e-04 & 5.3444e-03 & 3.1329e-04 & 2.8283e-03 & 2.2791e-04 \\
2.3500 & 4.8852e-03 & 2.9953e-04 & 2.2222e-03 & 2.0202e-04 & 4.1873e-03 & 2.7731e-04 & 1.7264e-03 & 1.7806e-04 \\
2.4500 & 3.6915e-03 & 2.6038e-04 & 1.3407e-03 & 1.5691e-04 & 3.1221e-03 & 2.3946e-04 & 1.0836e-03 & 1.4107e-04 \\
2.5500 & 2.9385e-03 & 2.3231e-04 & 1.1019e-03 & 1.4226e-04 & 2.2957e-03 & 2.0533e-04 & 9.3664e-04 & 1.3116e-04 \\
2.6500 & 1.8182e-03 & 1.8273e-04 & 6.7952e-04 & 1.1171e-04 & 1.4876e-03 & 1.6529e-04 & 8.6318e-04 & 1.2591e-04 \\
2.7500 & 1.2672e-03 & 1.5256e-04 & 6.4279e-04 & 1.0865e-04 & 1.0101e-03 & 1.3620e-04 & 7.1625e-04 & 1.1469e-04 \\
2.8500 & 1.1938e-03 & 1.4807e-04 & 3.1221e-04 & 7.5723e-05 & 7.1625e-04 & 1.1469e-04 & 2.3875e-04 & 6.6218e-05 \\
2.9500 & 7.5298e-04 & 1.1760e-04 & 4.2241e-04 & 8.8078e-05 & 6.6116e-04 & 1.1019e-04 & 2.0202e-04 & 6.0911e-05 \\

\end{longtable}

\newpage 
\subsection{Numerical data used in Fig.~\ref{fig:plot_npart_ratio}, Fig.~\ref{fig:plot_t_deltat_np} and p+p reference points for $\sqrt{s_{NN}}=8.8, 11.9, 16.8$~GeV}

\begin{sidewaystable}
\centering

\caption{Numerical values used for the centrality dependence in Ar+Sc collisions at $\sqrt{s_{NN}}=8.8$~GeV. The table lists the number of participants $N_\mathrm{part}$, the yields, corresponding statistical uncertainties, and the mean transverse momenta $\langle p_T\rangle$ for $K^+$, $K^-$, $K^{*0}$, and $\bar{K}^{*0}$. These values are used to construct the ratios shown in Fig.~\ref{fig:plot_npart_ratio} and the lifetime estimates shown in Fig.~\ref{fig:plot_t_deltat_np}.}
\label{tab:npart_ecm8p8}

\begin{tabular}{c c c c c c c c c c c c c}
\toprule
$N_\mathrm{part}$ & $K^+$ & $\Delta K^+$ & $\langle p_T\rangle^{K^+}$ & $K^-$ & $\Delta K^-$ & $\langle p_T\rangle^{K^-}$ & $K^{*0}$ & $\Delta K^{*0}$ & $\langle p_T\rangle^{K^{*0}}$ & $\bar{K}^{*0}$ & $\Delta \bar{K}^{*0}$ & $\langle p_T\rangle^{\bar{K^{*0}}}$ \\
\midrule
0.0000  & 1.3921e-01 & 2.1978e-03 & 4.4160e-01 & 7.0021e-02 & 1.5587e-03 & 4.0461e-01 & 4.5350e-02 & 1.2544e-03 & 5.1902e-01 & 1.5059e-02 & 7.2285e-04 & 5.3396e-01 \\
5.0000  & 4.6309e-01 & 2.8475e-03 & 4.3888e-01 & 2.3163e-01 & 2.0139e-03 & 3.9926e-01 & 1.4463e-01 & 1.5913e-03 & 5.3497e-01 & 4.7906e-02 & 9.1586e-04 & 5.3340e-01 \\
10.0000 & 8.2386e-01 & 3.6576e-03 & 4.4845e-01 & 4.1637e-01 & 2.6002e-03 & 4.0899e-01 & 2.4480e-01 & 1.9938e-03 & 5.4582e-01 & 8.4765e-02 & 1.1732e-03 & 5.4885e-01 \\
15.0000 & 1.2237e+00 & 4.7935e-03 & 4.5058e-01 & 6.1331e-01 & 3.3935e-03 & 4.1257e-01 & 3.5618e-01 & 2.5861e-03 & 5.5155e-01 & 1.2498e-01 & 1.5319e-03 & 5.6205e-01 \\
20.0000 & 1.6438e+00 & 6.0746e-03 & 4.5272e-01 & 8.2539e-01 & 4.3046e-03 & 4.1683e-01 & 4.7303e-01 & 3.2587e-03 & 5.5870e-01 & 1.6680e-01 & 1.9351e-03 & 5.6654e-01 \\
25.0000 & 2.1149e+00 & 7.4557e-03 & 4.5565e-01 & 1.0622e+00 & 5.2839e-03 & 4.2035e-01 & 5.8487e-01 & 3.9208e-03 & 5.6532e-01 & 2.1038e-01 & 2.3515e-03 & 5.6883e-01 \\
30.0000 & 2.5990e+00 & 8.8477e-03 & 4.5796e-01 & 1.3118e+00 & 6.2860e-03 & 4.2244e-01 & 7.2934e-01 & 4.6870e-03 & 5.7274e-01 & 2.5190e-01 & 2.7545e-03 & 5.8185e-01 \\
35.0000 & 3.0941e+00 & 1.0188e-02 & 4.5861e-01 & 1.5646e+00 & 7.2446e-03 & 4.2498e-01 & 8.4257e-01 & 5.3165e-03 & 5.7966e-01 & 3.0054e-01 & 3.1752e-03 & 5.8804e-01 \\
40.0000 & 3.5903e+00 & 1.1597e-02 & 4.6259e-01 & 1.8012e+00 & 8.2143e-03 & 4.2813e-01 & 9.7340e-01 & 6.0386e-03 & 5.8486e-01 & 3.4349e-01 & 3.5871e-03 & 5.9563e-01 \\
45.0000 & 4.1601e+00 & 1.3055e-02 & 4.6279e-01 & 2.1007e+00 & 9.2771e-03 & 4.3161e-01 & 1.1111e+00 & 6.7469e-03 & 5.8809e-01 & 3.9563e-01 & 4.0260e-03 & 5.9900e-01 \\
50.0000 & 4.6849e+00 & 1.4447e-02 & 4.6422e-01 & 2.3592e+00 & 1.0252e-02 & 4.3415e-01 & 1.2387e+00 & 7.4287e-03 & 5.8985e-01 & 4.4117e-01 & 4.4333e-03 & 6.0277e-01 \\
55.0000 & 5.2701e+00 & 1.5965e-02 & 4.6495e-01 & 2.6850e+00 & 1.1396e-02 & 4.3495e-01 & 1.3924e+00 & 8.2063e-03 & 5.9590e-01 & 4.9744e-01 & 4.9050e-03 & 6.1522e-01 \\
60.0000 & 5.8375e+00 & 1.7199e-02 & 4.6583e-01 & 2.9717e+00 & 1.2271e-02 & 4.3781e-01 & 1.5284e+00 & 8.8005e-03 & 6.0023e-01 & 5.5260e-01 & 5.2917e-03 & 6.0981e-01 \\
65.0000 & 6.4533e+00 & 1.8920e-02 & 4.6636e-01 & 3.2923e+00 & 1.3514e-02 & 4.4024e-01 & 1.7003e+00 & 9.7118e-03 & 6.0202e-01 & 5.9328e-01 & 5.7368e-03 & 6.1662e-01 \\
70.0000 & 7.0598e+00 & 2.1825e-02 & 4.6677e-01 & 3.5921e+00 & 1.5568e-02 & 4.4406e-01 & 1.8399e+00 & 1.1142e-02 & 6.0560e-01 & 6.5576e-01 & 6.6517e-03 & 6.1932e-01 \\
75.0000 & 7.5534e+00 & 3.4574e-02 & 4.7058e-01 & 3.8307e+00 & 2.4621e-02 & 4.4620e-01 & 1.9494e+00 & 1.7564e-02 & 6.1207e-01 & 7.1103e-01 & 1.0608e-02 & 6.3629e-01 \\
80.0000 & 7.9581e+00 & 1.2603e-01 & 4.6459e-01 & 4.1357e+00 & 9.0857e-02 & 4.4128e-01 & 2.1238e+00 & 6.5108e-02 & 6.1242e-01 & 8.0240e-01 & 4.0020e-02 & 6.0105e-01 \\
\bottomrule
\end{tabular}
\end{sidewaystable}

\begin{sidewaystable}
\centering

\caption{Numerical values used for the centrality dependence in Ar+Sc collisions at $\sqrt{s_{NN}}=11.9$~GeV. The table lists the number of participants $N_\mathrm{part}$, the yields, corresponding statistical uncertainties, and the mean transverse momenta $\langle p_T\rangle$ for $K^+$, $K^-$, $K^{*0}$, and $\bar{K}^{*0}$. These values are used to construct the ratios shown in Fig.~\ref{fig:plot_npart_ratio} and the lifetime estimates shown in Fig.~\ref{fig:plot_t_deltat_np}.}
\label{tab:npart_ecm11p9}

\begin{tabular}{c c c c c c c c c c c c c}
\toprule
$N_\mathrm{part}$ & $K^+$ & $\Delta K^+$ & $\langle p_T\rangle^{K^+}$ & $K^-$ & $\Delta K^-$ & $\langle p_T\rangle^{K^-}$ & $K^{*0}$ & $\Delta K^{*0}$ & $\langle p_T\rangle^{K^{*0}}$ & $\bar{K}^{*0}$ & $\Delta \bar{K}^{*0}$ & $\langle p_T\rangle^{\bar{K^{*0}}}$ \\
\midrule
0.0000  & 2.0114e-01 & 2.6416e-03 & 4.4179e-01 & 1.2073e-01 & 2.0465e-03 & 4.2208e-01 & 6.6435e-02 & 1.5182e-03 & 5.1710e-01 & 2.9870e-02 & 1.0180e-03 & 5.3191e-01 \\
5.0000  & 6.5343e-01 & 3.3842e-03 & 4.4442e-01 & 3.8330e-01 & 2.5920e-03 & 4.1965e-01 & 2.1345e-01 & 1.9342e-03 & 5.4015e-01 & 9.6032e-02 & 1.2974e-03 & 5.4672e-01 \\
10.0000 & 1.1531e+00 & 4.3198e-03 & 4.4884e-01 & 6.7355e-01 & 3.3016e-03 & 4.2390e-01 & 3.6501e-01 & 2.4305e-03 & 5.4417e-01 & 1.6475e-01 & 1.6328e-03 & 5.5744e-01 \\
15.0000 & 1.6805e+00 & 5.6175e-03 & 4.5260e-01 & 9.7765e-01 & 4.2847e-03 & 4.2420e-01 & 5.1966e-01 & 3.1238e-03 & 5.5186e-01 & 2.3211e-01 & 2.0877e-03 & 5.6306e-01 \\
20.0000 & 2.2706e+00 & 7.1499e-03 & 4.5547e-01 & 1.3156e+00 & 5.4424e-03 & 4.2792e-01 & 6.9233e-01 & 3.9480e-03 & 5.5983e-01 & 3.0605e-01 & 2.6250e-03 & 5.6754e-01 \\
25.0000 & 2.8909e+00 & 8.7331e-03 & 4.5698e-01 & 1.6732e+00 & 6.6440e-03 & 4.3234e-01 & 8.5585e-01 & 4.7517e-03 & 5.6764e-01 & 3.8879e-01 & 3.2026e-03 & 5.7440e-01 \\
30.0000 & 3.5174e+00 & 1.0259e-02 & 4.5886e-01 & 2.0427e+00 & 7.8178e-03 & 4.3409e-01 & 1.0286e+00 & 5.5476e-03 & 5.7445e-01 & 4.5440e-01 & 3.6873e-03 & 5.8249e-01 \\
35.0000 & 4.2014e+00 & 1.1903e-02 & 4.6021e-01 & 2.4194e+00 & 9.0324e-03 & 4.3591e-01 & 1.2020e+00 & 6.3666e-03 & 5.7826e-01 & 5.4257e-01 & 4.2774e-03 & 5.8661e-01 \\
40.0000 & 4.8256e+00 & 1.3418e-02 & 4.6117e-01 & 2.8159e+00 & 1.0250e-02 & 4.3852e-01 & 1.3772e+00 & 7.1682e-03 & 5.8042e-01 & 6.1656e-01 & 4.7963e-03 & 5.9102e-01 \\
45.0000 & 5.5717e+00 & 1.5115e-02 & 4.6299e-01 & 3.2390e+00 & 1.1524e-02 & 4.3944e-01 & 1.5679e+00 & 8.0182e-03 & 5.8485e-01 & 7.0834e-01 & 5.3893e-03 & 5.9590e-01 \\
50.0000 & 6.2661e+00 & 1.6691e-02 & 4.6435e-01 & 3.6439e+00 & 1.2728e-02 & 4.4224e-01 & 1.7414e+00 & 8.7990e-03 & 5.9133e-01 & 7.8179e-01 & 5.8956e-03 & 6.0517e-01 \\
55.0000 & 7.0469e+00 & 1.8430e-02 & 4.6577e-01 & 4.1040e+00 & 1.4065e-02 & 4.4691e-01 & 1.9467e+00 & 9.6868e-03 & 5.9158e-01 & 8.6002e-01 & 6.4385e-03 & 6.1038e-01 \\
60.0000 & 7.8047e+00 & 2.0069e-02 & 4.6725e-01 & 4.5492e+00 & 1.5322e-02 & 4.4636e-01 & 2.1425e+00 & 1.0515e-02 & 5.9680e-01 & 9.4974e-01 & 7.0008e-03 & 6.1234e-01 \\
65.0000 & 8.5924e+00 & 2.1712e-02 & 4.6859e-01 & 5.0166e+00 & 1.6590e-02 & 4.4760e-01 & 2.3469e+00 & 1.1347e-02 & 6.0269e-01 & 1.0391e+00 & 7.5503e-03 & 6.1784e-01 \\
70.0000 & 9.3985e+00 & 2.5230e-02 & 4.6879e-01 & 5.5048e+00 & 1.9309e-02 & 4.5071e-01 & 2.5493e+00 & 1.3140e-02 & 6.0669e-01 & 1.1524e+00 & 8.8345e-03 & 6.2038e-01 \\
75.0000 & 1.0075e+01 & 3.9912e-02 & 4.6828e-01 & 5.8854e+00 & 3.0504e-02 & 4.5149e-01 & 2.6783e+00 & 2.0578e-02 & 6.0345e-01 & 1.2030e+00 & 1.3791e-02 & 6.2335e-01 \\
80.0000 & 1.0844e+01 & 1.4004e-01 & 4.6767e-01 & 6.2604e+00 & 1.0640e-01 & 4.6074e-01 & 2.9620e+00 & 7.3187e-02 & 6.0432e-01 & 1.2857e+00 & 4.8218e-02 & 6.4511e-01 \\
\bottomrule
\end{tabular}
\end{sidewaystable}

\begin{sidewaystable}
\centering

\caption{Numerical values used for the centrality dependence in Ar+Sc collisions at $\sqrt{s_{NN}}=16.8$~GeV. The table lists the number of participants $N_\mathrm{part}$, the yields, corresponding statistical uncertainties, and the mean transverse momenta $\langle p_T\rangle$ for $K^+$, $K^-$, $K^{*0}$, and $\bar{K}^{*0}$. These values are used to construct the ratios shown in Fig.~\ref{fig:plot_npart_ratio} and the lifetime estimates shown in Fig.~\ref{fig:plot_t_deltat_np}.}
\label{tab:npart_ecm16p8}

\begin{tabular}{c c c c c c c c c c c c c}
\toprule
$N_\mathrm{part}$ & $K^+$ & $\Delta K^+$ & $\langle p_T\rangle^{K^+}$ & $K^-$ & $\Delta K^-$ & $\langle p_T\rangle^{K^-}$ & $K^{*0}$ & $\Delta K^{*0}$ & $\langle p_T\rangle^{K^{*0}}$ & $\bar{K}^{*0}$ & $\Delta \bar{K}^{*0}$ & $\langle p_T\rangle^{\bar{K^{*0}}}$ \\
\midrule
0.0000  & 2.8727e-01 & 3.1687e-03 & 4.4915e-01 & 1.8759e-01 & 2.5605e-03 & 4.2987e-01 & 1.0286e-01 & 1.8961e-03 & 5.2965e-01 & 5.5853e-02 & 1.3972e-03 & 5.4278e-01 \\
5.0000  & 9.0784e-01 & 3.9754e-03 & 4.4762e-01 & 5.9346e-01 & 3.2142e-03 & 4.2976e-01 & 3.1384e-01 & 2.3374e-03 & 5.3988e-01 & 1.7158e-01 & 1.7282e-03 & 5.4788e-01 \\
10.0000 & 1.5700e+00 & 5.0406e-03 & 4.5270e-01 & 1.0186e+00 & 4.0601e-03 & 4.3386e-01 & 5.2130e-01 & 2.9046e-03 & 5.5062e-01 & 2.8362e-01 & 2.1425e-03 & 5.5739e-01 \\
15.0000 & 2.2991e+00 & 6.5811e-03 & 4.5411e-01 & 1.4923e+00 & 5.3021e-03 & 4.3462e-01 & 7.4410e-01 & 3.7440e-03 & 5.5500e-01 & 4.0762e-01 & 2.7711e-03 & 5.6324e-01 \\
20.0000 & 3.0691e+00 & 8.3080e-03 & 4.5604e-01 & 1.9889e+00 & 6.6880e-03 & 4.3653e-01 & 9.7731e-01 & 4.6882e-03 & 5.5955e-01 & 5.2156e-01 & 3.4248e-03 & 5.6823e-01 \\
25.0000 & 3.8937e+00 & 1.0146e-02 & 4.5759e-01 & 2.5261e+00 & 8.1725e-03 & 4.3907e-01 & 1.2024e+00 & 5.6384e-03 & 5.6512e-01 & 6.5034e-01 & 4.1466e-03 & 5.7291e-01 \\
30.0000 & 4.7274e+00 & 1.1875e-02 & 4.5983e-01 & 3.0404e+00 & 9.5233e-03 & 4.4000e-01 & 1.4400e+00 & 6.5538e-03 & 5.7298e-01 & 7.7130e-01 & 4.7966e-03 & 5.8122e-01 \\
35.0000 & 5.6016e+00 & 1.3766e-02 & 4.6079e-01 & 3.6351e+00 & 1.1089e-02 & 4.4230e-01 & 1.6825e+00 & 7.5443e-03 & 5.7443e-01 & 9.0958e-01 & 5.5470e-03 & 5.8777e-01 \\
40.0000 & 6.4910e+00 & 1.5528e-02 & 4.6314e-01 & 4.2050e+00 & 1.2498e-02 & 4.4628e-01 & 1.9211e+00 & 8.4478e-03 & 5.8187e-01 & 1.0297e+00 & 6.1847e-03 & 5.9041e-01 \\
45.0000 & 7.4293e+00 & 1.7329e-02 & 4.6323e-01 & 4.8211e+00 & 1.3960e-02 & 4.4810e-01 & 2.1773e+00 & 9.3813e-03 & 5.8567e-01 & 1.1608e+00 & 6.8498e-03 & 5.9778e-01 \\
50.0000 & 8.3431e+00 & 1.9277e-02 & 4.6551e-01 & 5.4239e+00 & 1.5543e-02 & 4.4990e-01 & 2.4357e+00 & 1.0416e-02 & 5.8985e-01 & 1.3014e+00 & 7.6136e-03 & 6.0465e-01 \\
55.0000 & 9.3590e+00 & 2.1245e-02 & 4.6726e-01 & 6.0634e+00 & 1.7100e-02 & 4.5102e-01 & 2.6902e+00 & 1.1390e-02 & 5.9429e-01 & 1.4227e+00 & 8.2831e-03 & 6.0797e-01 \\
60.0000 & 1.0400e+01 & 2.3177e-02 & 4.6772e-01 & 6.7658e+00 & 1.8694e-02 & 4.5457e-01 & 2.9895e+00 & 1.2426e-02 & 5.9672e-01 & 1.6024e+00 & 9.0976e-03 & 6.1199e-01 \\
65.0000 & 1.1406e+01 & 2.5168e-02 & 4.7015e-01 & 7.4261e+00 & 2.0308e-02 & 4.5483e-01 & 3.2446e+00 & 1.3423e-02 & 6.0117e-01 & 1.7218e+00 & 9.7784e-03 & 6.1401e-01 \\
70.0000 & 1.2478e+01 & 2.9175e-02 & 4.7043e-01 & 8.1133e+00 & 2.3526e-02 & 4.5710e-01 & 3.5328e+00 & 1.5524e-02 & 6.0353e-01 & 1.8771e+00 & 1.1316e-02 & 6.1728e-01 \\
75.0000 & 1.3348e+01 & 4.5946e-02 & 4.6992e-01 & 8.6448e+00 & 3.6976e-02 & 4.6011e-01 & 3.7322e+00 & 2.4295e-02 & 6.0245e-01 & 2.0098e+00 & 1.7829e-02 & 6.1900e-01 \\
80.0000 & 1.3954e+01 & 1.6656e-01 & 4.7323e-01 & 9.2425e+00 & 1.3555e-01 & 4.6141e-01 & 4.0417e+00 & 8.9640e-02 & 6.1933e-01 & 2.1610e+00 & 6.5546e-02 & 6.2207e-01 \\
\bottomrule
\end{tabular}
\end{sidewaystable}

\begin{sidewaystable}
\centering
\caption{Numerical values used for the p+p reference points in Fig.~\ref{fig:plot_npart_ratio} and Fig.~\ref{fig:plot_t_deltat_np}. For p+p collisions, only the physical reference point at $N_\mathrm{part}=2$ is listed.}
\label{tab:pp_reference_points}
\begin{tabular}{c c c c c c c c c c c c c c}
\toprule
$\sqrt{s_{NN}}$ & $N_\mathrm{part}$ & $K^+$ & $\Delta K^+$ & $\langle p_T\rangle^{K^+}$ & $K^-$ & $\Delta K^-$ & $\langle p_T\rangle^{K^-}$ & $K^{*0}$ & $\Delta K^{*0}$ & $\langle p_T\rangle^{K^{*0}}$ & $\bar{K}^{*0}$ & $\Delta \bar{K}^{*0}$ & $\langle p_T\rangle^{\bar{K^{*0}}}$ \\
\midrule
8.8  & 0.0000 & 8.5539e-02 & 9.2487e-05 & 4.2294e-01 & 3.7075e-02 & 6.0889e-05 & 3.9887e-01 & 2.7128e-02 & 5.2084e-05 & 4.9295e-01 & 9.0603e-03 & 3.0100e-05 & 5.1344e-01 \\
12.3 & 0.0000 & 1.4381e-01 & 1.1992e-04 & 4.3525e-01 & 8.0371e-02 & 8.9650e-05 & 4.2322e-01 & 4.9853e-02 & 7.0607e-05 & 5.1705e-01 & 2.3574e-02 & 4.8553e-05 & 5.3677e-01 \\
17.3 & 0.0000 & 2.1250e-01 & 1.4577e-04 & 4.4203e-01 & 1.3654e-01 & 1.1685e-04 & 4.3394e-01 & 7.7345e-02 & 8.7946e-05 & 5.2647e-01 & 4.4690e-02 & 6.6850e-05 & 5.4196e-01 \\
\bottomrule
\end{tabular}
\end{sidewaystable}


\newpage
\begin{thebibliography}{99}

\bibitem{Sorensen:2023zkk}
A.~Sorensen, K.~Agarwal, K.~W.~Brown, Z.~Chaj{\k{e}}cki, P.~Danielewicz, C.~Drischler, S.~Gandolfi, J.~W.~Holt, M.~Kaminski and C.~M.~Ko, \textit{et al.}
Prog. Part. Nucl. Phys. \textbf{134}, 104080 (2024)
doi:10.1016/j.ppnp.2023.104080
[arXiv:2301.13253 [nucl-th]].

\bibitem{HanburyBrown:1954amm}
R.~Hanbury Brown and R.~Q.~Twiss,
Phil. Mag. Ser. 7 \textbf{45}, 663-682 (1954)
doi:10.1080/14786440708520475

\bibitem{Goldhaber:1960sf}
G.~Goldhaber, S.~Goldhaber, W.~Y.~Lee and A.~Pais,
Phys. Rev. \textbf{120}, 300-312 (1960)
doi:10.1103/PhysRev.120.300

\bibitem{Bauer:1992ffu}
W.~Bauer, C.~K.~Gelbke and S.~Pratt,
Ann. Rev. Nucl. Part. Sci. \textbf{42}, 77-100 (1992)
doi:10.1146/annurev.ns.42.120192.000453

\bibitem{Pratt:1986cc}
S.~Pratt,
Phys. Rev. D \textbf{33}, 1314-1327 (1986)
doi:10.1103/PhysRevD.33.1314

\bibitem{STAR:2014shf}
L.~Adamczyk \textit{et al.} [STAR],
Phys. Rev. C \textbf{92}, no.1, 014904 (2015)
doi:10.1103/PhysRevC.92.014904
[arXiv:1403.4972 [nucl-ex]].

\bibitem{PHENIX:2014dmi}
N.~N.~Ajitanand \textit{et al.} [PHENIX],
Nucl. Phys. A \textbf{931}, 1082-1087 (2014)
doi:10.1016/j.nuclphysa.2014.08.054
[arXiv:1404.5291 [nucl-ex]].

\bibitem{Armesto:2015ioy}
N.~Armesto and E.~Scomparin,
Eur. Phys. J. Plus \textbf{131}, no.3, 52 (2016)
doi:10.1140/epjp/i2016-16052-4
[arXiv:1511.02151 [nucl-ex]].

\bibitem{HADES:2019lek}
J.~Adamczewski-Musch \textit{et al.} [HADES],
Eur. Phys. J. A \textbf{56}, no.5, 140 (2020)
doi:10.1140/epja/s10050-020-00116-w
[arXiv:1910.07885 [nucl-ex]].

\bibitem{Li:2022iil}
P.~Li, J.~Steinheimer, T.~Reichert, A.~Kittiratpattana, M.~Bleicher and Q.~Li,
Sci. China Phys. Mech. Astron. \textbf{66}, no.3, 232011 (2023)
doi:10.1007/s11433-022-2041-8
[arXiv:2209.01413 [nucl-th]].

\bibitem{Torrieri:2001ue}
G.~Torrieri and J.~Rafelski,
Phys. Lett. B \textbf{509}, 239-245 (2001)
doi:10.1016/S0370-2693(01)00492-0
[arXiv:hep-ph/0103149 [hep-ph]].

\bibitem{Rafelski:2001hp}
J.~Rafelski, J.~Letessier and G.~Torrieri,
Phys. Rev. C \textbf{64}, 054907 (2001)
[erratum: Phys. Rev. C \textbf{65}, 069902 (2002)]
doi:10.1103/PhysRevC.64.054907
[arXiv:nucl-th/0104042 [nucl-th]].

\bibitem{Bleicher:2002dm}
M.~Bleicher and J.~Aichelin,
Phys. Lett. B \textbf{530}, 81-87 (2002)
doi:10.1016/S0370-2693(02)01334-5
[arXiv:hep-ph/0201123 [hep-ph]].

\bibitem{Markert:2002rw}
C.~Markert, G.~Torrieri and J.~Rafelski,
AIP Conf. Proc. \textbf{631}, no.1, 533 (2002)
doi:10.1063/1.1513698
[arXiv:hep-ph/0206260 [hep-ph]].

\bibitem{Knospe:2015rja}
A.~G.~Knospe [ALICE],
J. Phys. Conf. Ser. \textbf{612}, no.1, 012064 (2015)
doi:10.1088/1742-6596/612/1/012064
[arXiv:1501.03797 [nucl-ex]].

\bibitem{Knospe:2015nva}
A.~G.~Knospe, C.~Markert, K.~Werner, J.~Steinheimer and M.~Bleicher,
Phys. Rev. C \textbf{93}, no.1, 014911 (2016)
doi:10.1103/PhysRevC.93.014911
[arXiv:1509.07895 [nucl-th]].

\bibitem{Ilner:2016xqr}
A.~Ilner, D.~Cabrera, C.~Markert and E.~Bratkovskaya,
Phys. Rev. C \textbf{95}, no.1, 014903 (2017)
doi:10.1103/PhysRevC.95.014903
[arXiv:1609.02778 [hep-ph]].

\bibitem{Oliinychenko:2021enj}
D.~Oliinychenko and C.~Shen,
[arXiv:2105.07539 [hep-ph]].

\bibitem{Sahoo:2023rko}
A.~K.~Sahoo, M.~Nasim and S.~Singha,
Phys. Rev. C \textbf{108}, no.4, 044904 (2023)
doi:10.1103/PhysRevC.108.044904
[arXiv:2302.08177 [nucl-th]].

\bibitem{Sahoo:2023dkv}
A.~K.~Sahoo, S.~Singha and M.~Nasim,
J. Phys. G \textbf{52}, no.1, 015101 (2025)
doi:10.1088/1361-6471/ad8768
[arXiv:2307.06661 [nucl-th]].

\bibitem{STAR:2004bgh}
J.~Adams \textit{et al.} [STAR],
Phys. Rev. C \textbf{71}, 064902 (2005)
doi:10.1103/PhysRevC.71.064902
[arXiv:nucl-ex/0412019 [nucl-ex]].

\bibitem{Markert:2005jv}
C.~Markert,
J. Phys. G \textbf{31}, S169-S178 (2005)
doi:10.1088/0954-3899/31/4/021
[arXiv:nucl-ex/0503013 [nucl-ex]].

\bibitem{STAR:2008twt}
B.~I.~Abelev \textit{et al.} [STAR],
Phys. Rev. C \textbf{78}, 044906 (2008)
doi:10.1103/PhysRevC.78.044906
[arXiv:0801.0450 [nucl-ex]].

\bibitem{HADES:2013sfy}
G.~Agakishiev \textit{et al.} [HADES],
Eur. Phys. J. A \textbf{49}, 34 (2013)
doi:10.1140/epja/i2013-13034-7

\bibitem{ALICE:2018ewo}
S.~Acharya \textit{et al.} [ALICE],
Phys. Rev. C \textbf{99}, 024905 (2019)
doi:10.1103/PhysRevC.99.024905
[arXiv:1805.04361 [nucl-ex]].

\bibitem{ALICE:2022zuc}
S.~Acharya \textit{et al.} [ALICE],
Eur. Phys. J. C \textbf{83}, no.5, 351 (2023)
doi:10.1140/epjc/s10052-023-11475-1
[arXiv:2205.13998 [nucl-ex]].

\bibitem{ALICE:2023ifn}
S.~Acharya \textit{et al.} [ALICE],
Phys. Rev. C \textbf{109}, no.4, 044902 (2024)
doi:10.1103/PhysRevC.109.044902
[arXiv:2308.16119 [nucl-ex]].

\bibitem{Knospe:2021jgt}
A.~G.~Knospe, C.~Markert, K.~Werner, J.~Steinheimer and M.~Bleicher,
Phys. Rev. C \textbf{104}, no.5, 054907 (2021)
doi:10.1103/PhysRevC.104.054907
[arXiv:2102.06797 [nucl-th]].

\bibitem{Chabane:2024crn}
A.~Chabane, L.~Engel, T.~Reichert, J.~Steinheimer and M.~Bleicher,
Phys. Rev. C \textbf{111}, no.3, 034913 (2025)
doi:10.1103/PhysRevC.111.034913
[arXiv:2409.08639 [hep-ph]].

\bibitem{Neidig:2025xgr}
T.~Neidig, A.~Kittiratpattana, T.~Reichert, A.~Chabane, C.~Greiner and M.~Bleicher,
Phys. Lett. B \textbf{862}, 139339 (2025)
doi:10.1016/j.physletb.2025.139339
[arXiv:2501.03893 [nucl-th]].

\bibitem{Kozlowski:2024cjw}
B.~Koz{\l}owski [NA61/SHINE],
PoS \textbf{ICHEP2024}, 609 (2025)
doi:10.22323/1.476.0609
[arXiv:2409.20229 [nucl-ex]].

\bibitem{STAR:2022sir}
M.~Abdallah \textit{et al.} [STAR],
Phys. Rev. C \textbf{107}, no.3, 034907 (2023)
doi:10.1103/PhysRevC.107.034907
[arXiv:2210.02909 [nucl-ex]].

\bibitem{Bass:1998ca}
S.~A.~Bass, M.~Belkacem, M.~Bleicher, M.~Brandstetter, L.~Bravina, C.~Ernst, L.~Gerland, M.~Hofmann, S.~Hofmann and J.~Konopka, \textit{et al.}
Prog. Part. Nucl. Phys. \textbf{41}, 255-369 (1998)
doi:10.1016/S0146-6410(98)00058-1
[arXiv:nucl-th/9803035 [nucl-th]].

\bibitem{Bleicher:1999xi}
M.~Bleicher, E.~Zabrodin, C.~Spieles, S.~A.~Bass, C.~Ernst, S.~Soff, L.~Bravina, M.~Belkacem, H.~Weber and H.~Stoecker, \textit{et al.}
J. Phys. G \textbf{25}, 1859-1896 (1999)
doi:10.1088/0954-3899/25/9/308
[arXiv:hep-ph/9909407 [hep-ph]].

\bibitem{Bleicher:2022kcu}
M.~Bleicher and E.~Bratkovskaya,
Prog. Part. Nucl. Phys. \textbf{122}, 103920 (2022)
doi:10.1016/j.ppnp.2021.103920

\bibitem{Aichelin:1986wa}
J.~Aichelin and H.~Stoecker,
Phys. Lett. B \textbf{176}, 14-19 (1986)
doi:10.1016/0370-2693(86)90916-0

\bibitem{Aichelin:1991xy}
J.~Aichelin,
Phys. Rept. \textbf{202}, 233-360 (1991)
doi:10.1016/0370-1573(91)90094-3

\bibitem{ParticleDataGroup:2024cfk}
S.~Navas \textit{et al.} [Particle Data Group],
Phys. Rev. D \textbf{110}, no.3, 030001 (2024)
doi:10.1103/PhysRevD.110.030001

\bibitem{Vogel:2009kg}
S.~Vogel, J.~Aichelin and M.~Bleicher,
Phys. Rev. C \textbf{82}, 014907 (2010)
doi:10.1103/PhysRevC.82.014907
[arXiv:0908.3811 [hep-ph]].

\bibitem{Steinheimer:2015sha}
J.~Steinheimer and M.~Bleicher,
J. Phys. G \textbf{43}, no.1, 015104 (2016)
doi:10.1088/0954-3899/43/1/015104
[arXiv:1503.07305 [nucl-th]].

\bibitem{Ilner:2017tab}
A.~Ilner, J.~Blair, D.~Cabrera, C.~Markert and E.~Bratkovskaya,
Phys. Rev. C \textbf{99}, no.2, 024914 (2019)
doi:10.1103/PhysRevC.99.024914
[arXiv:1707.00060 [hep-ph]].

\bibitem{Reichert:2019lny}
T.~Reichert, P.~Hillmann, A.~Limphirat, C.~Herold and M.~Bleicher,
J. Phys. G \textbf{46}, no.10, 105107 (2019)
doi:10.1088/1361-6471/ab34fa
[arXiv:1903.12032 [nucl-th]].

\bibitem{Neidig:2021bal}
T.~Neidig, K.~Gallmeister, C.~Greiner, M.~Bleicher and V.~Vovchenko,
Phys. Lett. B \textbf{827}, 136891 (2022)
doi:10.1016/j.physletb.2022.136891
[arXiv:2108.13151 [hep-ph]].

\end{thebibliography}
\end{document}